\documentclass[pdflatex,sn-nature,Numbered]{sn-jnl}

\usepackage{graphicx}%
\usepackage{multirow}%
\usepackage{amsmath,amssymb,amsfonts}%
\usepackage{amsthm}%
\usepackage{mathrsfs}%
\usepackage[title]{appendix}%
\usepackage{xcolor}%
\usepackage{textcomp}%
\usepackage{manyfoot}%
\usepackage{booktabs}%
\usepackage{algorithm}%
\usepackage{algorithmicx}%
\usepackage{algpseudocode}%
\usepackage{listings}%

\usepackage[detect-all=true]{siunitx}[=v2]
\usepackage{comment}
\usepackage{gensymb}
\usepackage[version=4]{mhchem}
\usepackage[T1]{fontenc}
\usepackage{lmodern}
\usepackage[scr=rsfso,cal=rsfso]{mathalfa} 
\begin{document}

\title[Article Title]{Cyclic jetting enables microbubble-mediated drug delivery}

\author*[1]{\fnm{Marco} \sur{Cattaneo}}\email{mcattaneo@ethz.ch}

\author[1]{\fnm{Giulia} \sur{Guerriero}}

\author[1]{\fnm{Gazendra} \sur{Shakya}}

\author[2]{\fnm{Lisa A.} \sur{Krattiger}}

\author[3,4]{\fnm{Lorenza} \sur{G. Paganella}}

\author[4,5]{\fnm{Maria L.} \sur{Narciso}}

\author*[1]{\fnm{Outi} \sur{Supponen}}\email{outis@ethz.ch}

\affil[1]{\orgdiv{Institute of Fluid Dynamics}, \orgname{ETH Zürich}, \orgaddress{\street{Sonneggstrasse 3}, \city{Z{\"u}rich}, \postcode{8092}, \country{Switzerland}}}

\affil[2]{\orgdiv{Department of Obstetrics}, \orgname{University Hospital Z{\"u}rich, University of Z{\"u}rich}, \orgaddress{\street{Schmelzbergstrasse 12}, \city{Z{\"u}rich}, \postcode{8091}, \country{Switzerland}}}

\affil[3]{\orgdiv{Institute of Energy and Process Engineering}, \orgname{ETH Zürich}, \orgaddress{\street{Leonhardstrasse 21}, \city{Z{\"u}rich}, \postcode{8092}, \country{Switzerland}}}

\affil[4]{\orgdiv{Institute for Mechanical Systems}, \orgname{ETH Zürich}, \orgaddress{\street{Leonhardstrasse 21}, \city{Z{\"u}rich}, \postcode{8092}, \country{Switzerland}}}

\affil[5]{\orgname{Swiss Federal Laboratories for Materials Science and Technology (EMPA)}, \orgaddress{\street{{\"U}berlandstrasse 129}, \city{Dübendorf}, \postcode{8600}, \country{Switzerland}}}

\abstract{
The pursuit of targeted therapies capable of overcoming biological barriers, including the tenacious blood-brain barrier, has spurred the investigation into stimuli-responsive microagents.
This approach could improve therapeutic efficacy, reduce undesirable side effects, and open avenues for treating previously incurable diseases.
Intravenously-administered ultrasound-responsive microbubbles are one of the most promising agents, having demonstrated potential in several clinical trials.
However, the mechanism by which microbubbles enhance drug absorption remains unclear.
Here, we reveal through unprecedented time-resolved side-view visualisations that single microbubbles, upon microsecond-long ultrasound driving, puncture the cell membrane and induce drug uptake via stable cyclic microjets.
Our theoretical models successfully reproduce the observed bubble and cell dynamic responses.
We find that cyclic jets arise from shape instabilities, warranting recognition as a novel class of jets in bubbles, distinct from classical inertial jets driven by pressure gradients.
We also establish a threshold for bubble radial expansion beyond which microjets form and facilitate cellular permeation.
Remarkably, these microjets occur at ultrasound pressures below \SI{100}{\kilo\pascal} due to their unique formation mechanism.
We show that the stress generated by microjetting surpasses all previously suggested mechanisms by at least an order of magnitude.
In summary, this work elucidates the physics behind microbubble-mediated targeted drug delivery and provides criteria for its effective yet safe application.
}


\maketitle

\newpage

Targeted drug delivery holds the potential to revolutionise healthcare by enhancing the precision of drug administration and thus minimising side effects \cite{Manzari2021TargetedMedicines}.
By using specially engineered vascular carriers, drugs are encapsulated, transported, and released at designated sites within the body \cite{Mitchell2021EngineeringDelivery,Ferguson2018KinaseAhead}.
Despite the ability of this approach to increase drug accumulation at targeted regions, its therapeutic efficacy is hindered by biological barriers, notably the endothelium and the blood-brain barrier (BBB), which tightly regulate the molecular passage between the bloodstream and tissues, thereby limiting the drug bioavailability \cite{Blanco2015PrinciplesDelivery}.

Ultrasound-responsive agents, such as phospholipid-coated microbubbles, offer solutions for enhancing specificity and overcoming biological barriers in drug delivery \cite{Kooiman2014AcousticDelivery,Stride2019NucleationDelivery,Kooiman2020Ultrasound-ResponsiveDelivery,Versluis2020UltrasoundReview,Shakya2024Ultrasound-responsiveDelivery}.
These agents, either co-administered with drugs in the systemic circulation or directly conjugated to them, are actuated with spatial precision by focused ultrasound systems  \cite{Meng2021TechnicalUltrasound}.
Ultrasound induces cyclic oscillations in the bubbles, generating mechanical stresses that temporarily open biological barriers, enabling drug delivery across them  \cite{Sheikov2004CellularMicrobubbles} (Fig.~\ref{fig:Fig1}a).

As of today, ultrasound-activated microbubbles are the only non-invasive, localised, and reversible method for opening the BBB and delivering drugs to the brain \cite{Konofagou2012OptimizationOpening}.
This technique holds promise for treating neurodegenerative disorders such as Alzheimer’s and Parkinson’s disease \cite{Lipsman2018BloodbrainUltrasound,Rezai2020NoninvasiveUltrasound,Gasca-Salas2021Blood-brainDementia}, brain tumors \cite{Carpentier2016ClinicalUltrasound,Idbaih2019SafetyGlioblastoma,Mainprize2019Blood-BrainStudy,Meng2021MR-guidedMetastases,Chen2021Neuronavigation-guidedTumors,Sonabend2023RepeatedTrial}, and amyotrophic lateral sclerosis in humans \cite{Abrahao2019First-in-humanUltrasound}.
Additionally, it shows potential for treating solid tumors \cite{Dimcevski2016ACancer,Eisenbrey2021US-triggeredTrial}, myocardial infarction \cite{Mathias2019SonothrombolysisIntervention}, and atherosclerosis \cite{Kopechek2019UltrasoundFunction}.

Despite promising clinical results, the physical mechanism by which microbubbles enhance biological barrier permeability remains unclear.
Proposed mechanisms include acoustic streaming \cite{Marmottant2003ControlledBubbles}, inertial jetting \cite{Prentice2005MembraneCavitation}, normal impact pressure \cite{Fan2012SpatiotemporallySonoporation}, and viscous shear stress \cite{Helfield2016BiophysicalSonoporation}.
The lack of consensus underscores the formidable challenge of directly observing bubble behaviour and correlating it with drug uptake, a crucial step for ensuring the safety of microbubble-mediated drug delivery.

Cell monolayers on rigid substrates are the primary \emph{in-vitro} platform for monitoring membrane integrity and drug uptake with ultrasound-driven microbubbles \cite{vanWamel2006VibratingSonoporation,Fan2012SpatiotemporallySonoporation,Helfield2016BiophysicalSonoporation, vanElburg2023DependenceStudy, Lajoinie2016}.
\emph{Ex-vivo} tissues \cite{Chen2011BloodCavitation,Bezer2023MicrobubbleMicrovessels} and \emph{in-vivo} embryos \cite{Anbarafshan2024InModel} have also been employed to investigate the dynamics of microbubbles within vascular structures, but the resulting drug uptake has not been analysed.
Current investigations using cell monolayers are constrained to a top-down view, which provides an incomplete  picture of the bubble-cell interplay.
In this study, we adopt a novel side-view perspective to explore the underlying physics. 
This approach presents significant challenges due to restricted optical access, which we address through a carefully designed experimental setup and test samples.
Our new viewpoint on the problem enables us to uncover new and key insights into the physics of microbubble-mediated drug delivery.
We anticipate our investigation to guide the future developments of this technology.

\newpage

\subsection*{Bubble jetting and sonoporation} 

To enable side-view visualisations of drug delivery facilitated by individual microbubbles, we culture human umbilical vein endothelial cells (HUVECs) on a plastic substrate that we position within a custom-designed test chamber.
The chamber features an acoustically transparent base and optically transparent sides.
We fill the chamber with a solution of phosphate-buffered saline (PBS), propidium iodide (PI), and phospholipid-coated microbubbles (\SIrange{1}{4}{\micro\meter} in radius).
Microbubbles adhere to cells through flotation.
The chamber is immersed in a water bath (Fig. 1b).
We use a single ultrasound pulse (frequency $f = \SI{1}{\mega\hertz}$, 20 cycles) with a ramp profile, directed from the bottom, to drive the microbubbles.
We capture the bubble response under varying ultrasound pressures using a custom-built side-view 200$\times$ microscope recording at 10 million frames per second and assess the cell membrane permeabilisation by observing the intracellular fluorescence of PI, which serves as a model drug (see Methods and Extended Data Fig.~\ref{fig:Setup} for details).

\begin{figure}[H] 
    \centering
        \includegraphics[width=\columnwidth]{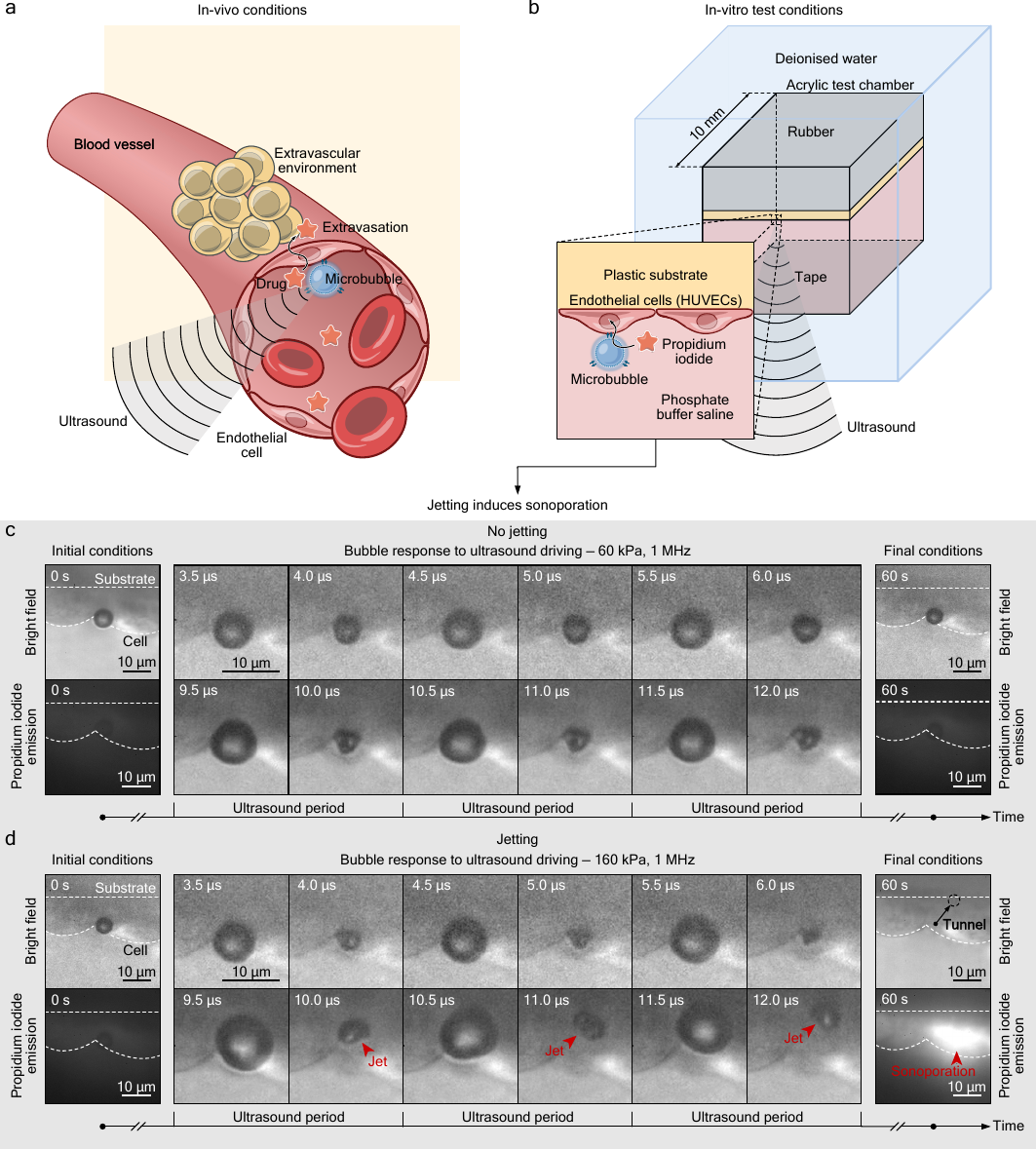}
    \caption{Targeted drug delivery mediated by ultrasound-responsive microbubbles. (a) Schematic illustrating \emph{in-vivo} extravascular drug delivery induced by the mechanical action exerted by an ultrasound-driven microbubble.
    (b) Schematic depicting the \emph{in-vitro} test model employed to study the bubble-cell dynamics and the corresponding intracellular drug uptake with a side-view perspective.
    See Methods and Extended Data Fig.~\ref{fig:Setup} for details about the experimental setup.
    (c-d) Response of a microbubble ($R_0 = \SI{3}{\micro\meter}$) to varying ultrasound pressure amplitudes and the corresponding model drug uptake.
    In (c), the applied ultrasound pressure induces spherical oscillations in the bubble, followed by asymmetric deformation ($p_{\rm a} = \SI{60}{\kilo\pascal}, \ f = \SI{1}{\mega\hertz}$).
    However, this does not result in cell membrane poration and drug uptake.
    In (d), a higher ultrasound pressure causes the bubble to develop cyclic piercing microjets directed towards the cell ($p_{\rm a} = \SI{160}{\kilo\pascal}, \ f = \SI{1}{\mega\hertz}$), resulting in cell membrane poration and drug uptake.
    }
    \label{fig:Fig1}
\end{figure}

At a mild ultrasound pressure ($p_{\rm a} = \SI{60}{\kilo\pascal}$), a single microbubble in contact with a cell undergoes alternating phases of expansion and compression while maintaining its spherical shape (Fig.~\ref{fig:Fig1}c, Supplementary Video 1, \SIrange{3.5}{6.0}{\micro\second}).
Over time, the microbubble starts to exhibit non-spherical compression phases, forming microjets (Fig.~\ref{fig:Fig1}c, Supplementary Video 1, \SIrange{9.5}{12.0}{\micro\second}).
These jets are aimed at the cell but lack sufficient momentum to pierce the bubble and impact the cell on the opposite side.
The cell membrane remains undamaged, as indicated by the absence of intracellular fluorescence (Fig.~\ref{fig:Fig1}c, 60 s).
At a higher ultrasound pressure ($p_{\rm a} = \SI{160}{\kilo\pascal}$), the same microbubble experiences a larger radial excursion and now develops piercing cyclic jets that hammer the cell at each compression phase (Fig.~\ref{fig:Fig1}d, Supplementary Video 2, \SIrange{9.5}{12.0}{\micro\second}).
This response results in cell membrane poration and PI uptake, evidenced by the intense fluorescent emission (Fig.~\ref{fig:Fig1}d, 60 s).
The process of mechanically opening the cell membrane using microbubbles and ultrasound is known as “sonoporation” \cite{Bao1997TransfectionVitro}.

Previous studies have shown that sonoporation can cause the opening of cell-cell contacts \cite{Beekers2020OpeningSonoporation} and, in some cases, lead to the formation of transendothelial tunnels —transcellular perforations that pierce both the apical and basal cell membranes—\cite{Helfield2020TransendothelialSonoporation,Beekers2022InternalizationTunnels} thereby facilitating transcellular drug delivery into the extravascular space.
While the side-view perspective does not allow visualisation of cell-cell openings, it does reveal that, in the current case of successful sonoporation, bubble activity not only deforms and punctures the apical cell membrane but also induces the formation of a transendothelial tunnel.
This is evident from the microbubble position after the ultrasound pulse, as it rests deflated against the plastic substrate at the basal membrane level (Fig.~\ref{fig:Fig1}d, 60 s).
However, tunnel formation does not occur in all sonoporation events.
If the cell membrane damage is insufficient to internalise the bubble, it returns to its initial position when the ultrasound pulse ends (Extended Data Fig.~\ref{fig:FigNoTunnel}, Supplementary Video 3). 
Nevertheless, whether through tunneling or reversible cell deformation, the bubble has the potential to reach and perforate the basal cell membrane, thus enabling drug transport into the extravascular space.

When ultrasound is applied without the presence of microbubbles, no sonoporation events occur at the tested ultrasound pressures (up to \SI{1}{\mega\pascal}), even after tens of repeated pulses.

\newpage

\newpage

\subsection*{Characterisation of bubble dynamics} 

\begin{figure}[H] 
    \centering
        \includegraphics[width=0.5\columnwidth]{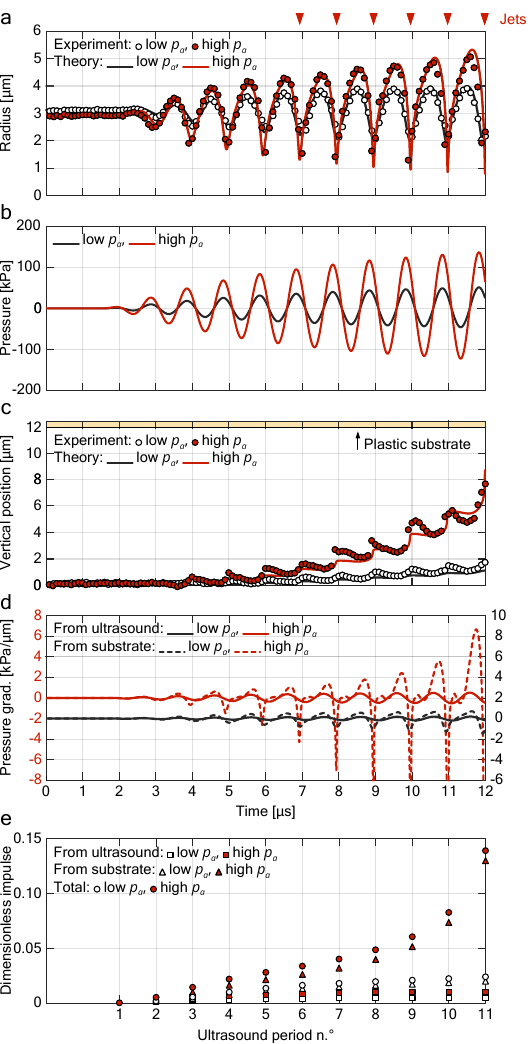}
    \caption{Characterisation of bubble dynamics.
    In the legend, ``low $p_{\rm a}$" and ``high $p_{\rm a}$" refer to Figs.~\ref{fig:Fig1}c and \ref{fig:Fig1}d, respectively.
    The red arrows at the top indicate when jets occur in the ``high $p_{\rm a}$" case.
    See Methods for details about the theoretical models.
    (a) Experimental and theoretical radial motion of the microbubble. 
    The uncertainty in the bubble radius measurement corresponds to half the pixel size (\SI{80}{\nano\meter}).
    (b) Ultrasound pulse driving the microbubble. 
    The pulse shape is recorded with a hydrophone.
    The pulse amplitude is inferred as the only fitting parameter from the corresponding radius-time curve in Fig.~\ref{fig:Fig2}a. 
    (c) Experimental and theoretical vertical position of the microbubble centroid. 
    The yellow area represents the plastic substrate. 
    The uncertainty in the bubble position measurement corresponds to half the pixel size (\SI{80}{\nano\meter}).
    (d) Time evolution of the pressure gradient contributions from the ultrasound pulse and plastic substrate at the bubble position.
    The curves for the two ultrasound pressure cases are vertically offset to enhance visibility.
    (e) Evolution of the dimensionless impulse contributions from the ultrasound, the plastic substrate and the sum thereof over ultrasound cycles. 
    }
    \label{fig:Fig2}
\end{figure}

The time evolution of the microbubble radius in the cases depicted in Fig.~\ref{fig:Fig1}c-d is extracted from video recordings and compared with theoretical predictions (Fig.~\ref{fig:Fig2}a).
The theoretical model incorporates a modified Rayleigh-Plesset equation for the liquid phase \cite{Brenner2002Single-bubbleSonoluminescence}, the Zhou thermal model for the gaseous phase \cite{Zhou2021ModelingBubble}, and the Marmottant model for the phospholipid coating \cite{Marmottant2005ARupture} (see Methods for details).
Despite assuming spherical symmetry, the model agrees well with observations, as the bubble remains mostly spherical due to the mild constriction of the soft cellular monolayer, deviating only during the final instants of the compression phase to form jets (Fig.~\ref{fig:Fig2}b).
The ultrasound pressure is experimentally measured using a hydrophone without the presence of the test chamber.
Its amplitude is then adjusted by applying an amplification factor obtained through a fitting procedure, matching experimental and theoretical radius-time curves, to account for the variable sound absorption by nearby bubbles within the test chamber, as well as the acoustic reflections at the bubble’s location (see Methods for details).

The time evolution of the vertical position of the microbubble centroid is also extracted from video recordings and compared with theoretical predictions (Fig.~\ref{fig:Fig2}c).
As the bubble displaces, it compresses the adjacent cell.
For the first, lower-amplitude ultrasound pulse, cell deformation is entirely reversible, as the bubble returns to its original position by the onset of the second pulse.
Conversely, the second, higher-amplitude pulse causes the cell to exceed its ultimate compressive strain, creating a transendothelial tunnel.
The bubble displacement is modelled through the force balance between the bubble inertia, the hydrodynamic forces, the cell resistive force, and the acoustic driving forces (see Methods for details).
The acoustic driving forces include the primary Bjerknes force from the ultrasound and the secondary Bjerknes force from the rigid plastic substrate, viewed as a virtual bubble emitting a secondary sound field \cite{Mettin1997BjerknesField}.
The cell layer contribution to the secondary Bjerknes force is negligible due to its softness, as is the buoyancy force due to the brief dynamics duration.
Given the overall good agreement with experiments, this theoretical model, without free parameters, could be useful to effectively simulate the deformation of living tissue caused by ultrasound-driven microbubbles.
Minor discrepancies observed during bubble compression phases can be attributed to the bubble collapse asymmetry, which is not accounted for in the spherically symmetric model.

In sonoporation studies, rigid plastic substrates are commonly used for growing endothelial cell monolayers because they significantly enhance cell proliferation, leading to a more uniform monolayer compared to soft substrates \cite{Yeh2012Matrix9}.
However, the influence on bubble dynamics of a rigid plate beneath a soft material layer remains underexplored.
Existing research on bubble dynamics involving such composite substrates is limited to numerical simulations conducted at high ultrasound pressures \cite{Curtiss2013UltrasonicLayer}.
We assess this gap by comparing the effect of the plastic substrate against that of ultrasound, which is always present and serves as a baseline.
By considering the pressure gradient from the ultrasound pulse and the pressure gradient from the sound field generated by the rigid backing plate (Fig.~\ref{fig:Fig2}d), we compute the corresponding dimensionless impulses that these pressure gradients induce in the fluid during each ultrasound period.
We extend the applicability of the dimensionless impulse, previously limited to collapsing vapour bubbles, to include bubbles subjected to a generic driving pressure (see Methods for details).
The impulse imparted by the plastic substrate surpasses that induced by ultrasound by a factor increasing from 3 to 15 as the bubble approaches the substrate (Fig.~\ref{fig:Fig2}e).
This suggests a potential influence on jet formation. 
To clarify this, we verify whether the jetting phenomenon persists when microbubbles interact with a thick soft substrate.

\newpage

\newpage

\subsection*{The nature of the jets} 

\begin{figure}[!ht] 
    \centering
        \includegraphics[width=\columnwidth]{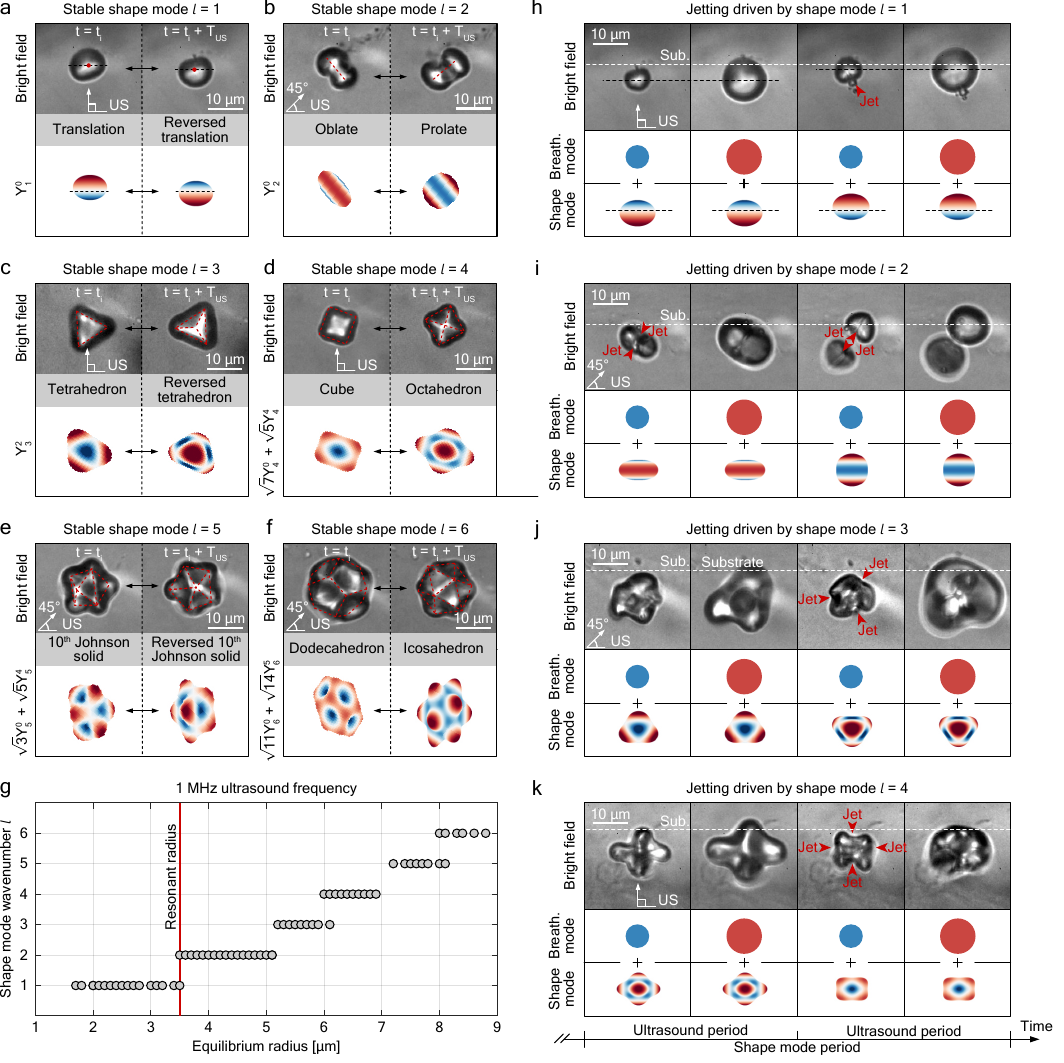}
    \caption{Shape modes and jet formation of single microbubbles in contact with a PEG substrate.
    (a-f) Shape modes with angular wavenumbers $l$ ranging from 1 up to 6.
    The top panels display bright-field snapshots from two consecutive ultrasound cycles, at a generic time instant $t = t_i$ and one ultrasound period $T_{\rm US}=\SI{1}{\micro\second}$ later, at $t = t_i + T_{\rm US}$. These images illustrate the cyclic transition between a geometric pattern and its dual during the shape mode oscillation.
    Odd wavenumber shape mode patterns are self-dual.
    The red dashed sketches show the polyhedral representation of shape mode patterns.
    On bottom, the combination of spherical harmonics $Y_l^m$ representing the shape mode patterns.        
    Red regions denote outward deformation, while blue regions indicate inward deformation.
    (g) Experimentally observed shape modes angular wavenumber $l$ as function of the equilibrium bubble radius. 
    The uncertainty in the bubble radius measurement corresponds to half the pixel size (\SI{80}{\nano\meter}).
    (h-k) Jets driven by shape modes with angular wavenumbers $l$ ranging from 1 up to 4.
    On top, bright field images depicting the jets generated by the shape pattern or by its dual.
    On bottom, bubble shape decomposition in breathing mode (spherical oscillation) and shape mode.
    Red regions denote outward deformation, while blue regions indicate inward deformation.
    Jets manifest during compression phases in the sunken regions of the shape mode. 
    }
    \label{fig:Fig3}
\end{figure}

We employ as a thick soft substrate a polyethylene glycol (PEG) hydrogel slab with a compression elastic modulus of $E\approx\SI{0.6}{\kilo\pascal}$ to mimic the softness of brain tissue \cite{Budday2015MechanicalIndentation} (see Methods for details).
The optical properties of this test model are superior to those of the cellular substrate and therefore enable for a more comprehensive investigation of bubble dynamics.
Upon ultrasound driving, we observe that the radial oscillation of the bubble destabilises its interface once a threshold amplitude is exceeded, leading to the formation of a stable standing wave pattern on the bubble surface, with a frequency half that of the ultrasound driving.
This phenomenon is driven by the Faraday instability \cite{Faraday1831XVII.Surfaces}, which causes the appearance of half-harmonic patterns, known as shape modes, on oscillating density interfaces.
We identify shape modes with angular wavenumbers $l$ ranging from 1 to 6 (Fig.~\ref{fig:Fig3}a-f).
The wavenumber increases with bubble size (Fig.~\ref{fig:Fig3}g).
A shape mode of wavenumber $l$ can be expressed using spherical harmonics $Y_l^m$ ($0 \le m \le l$), where $l$ and $m$ denote its degree and order, respectively.
The linear stability of the bubble interface is independent of the spherical harmonic order $m$ \cite{Rayleigh1879VI.Jets,Lamb1932Hydrodynamics}.
Consequently, each of the $l + 1$ possible $m$ values is equally probable from a linear standpoint.
The pattern that ultimately forms is thus a consequence of nonlinear effects.
Specifically, for $l = 1$, we observe the bubble undergoing an alternating rigid-body motion (Fig.~\ref{fig:Fig3}a).
For $l = 2$, the shape mode oscillates between oblate and prolate shapes (Fig.~\ref{fig:Fig3}b).
For higher $l$, the shape mode alternates between polyhedral patterns and their duals, where the vertices of one correspond to the faces of the other and viceversa (Fig.~\ref{fig:Fig3}c-f).
Previous studies have already observed shape modes on bubbles \cite{Guedra2016ExperimentalMicrobubbles,Guedra2017DynamicsThreshold}—including coated microbubbles \cite{Zhao2005AsymmetricAgents,Dollet2008NonsphericalMicrobubbles,Poulichet2017ShapeExpulsion}—but they were limited to reporting only the degree of the shape mode.
In contrast, as far as we know, we reveal for the first time the full three-dimensional pattern of these shape oscillations and identify the combination of spherical harmonics that describe them (see Fig.~\ref{fig:Fig3}a-f).
Our observations confirm past theoretical \cite{Busse1975PatternsShells,Busse1982Patterns2,Riahi1984NonlinearShell,Chossat1991Steady-State03-Symmetry} and numerical predictions \cite{Ebo-Adou2019FaradaySimulation} concerning the dominant three-dimensional patterns of the Faraday instability on a spherical interface.

When the shape mode amplitude is sufficiently large, the shape lobes fold in rapidly during bubble compression and shape reversal, generating cyclic microjets (Fig.~\ref{fig:Fig3}h-k).
We find that the number and orientation of the ejected jets depend on the specific shape mode.
The $l = 1$ mode generates alternately directed single jets (Fig.~\ref{fig:Fig3}h), while the $l = 2$ mode produces pairs of jets that alternately converge and diverge. (Fig.~\ref{fig:Fig3}i).
Higher wavenumbers $l$ result in multiple jets that tend to match the number of faces of the polyhedral pattern (Fig.~\ref{fig:Fig3}j-k).
Shape mode-induced cyclic jets on bubbles have likely already been observed in previous studies, which documented the occurrence of repeated jets resulting from surface deformation at driving frequencies spanning hertz \cite{Crum1979SurfaceBubbles}, kilohertz \cite{Prabowo2011SurfaceBubbles}, and even megahertz \cite{Vos2011} ranges.
This jetting phenomenon can be considered as the spherical analog to the jets formed by Faraday waves on vertically-vibrating liquid baths \citep{Longuet-Higgins1983BubblesSurface, Zeff2000SingularitySurface}.
Such jets emerge when the depressions created by the Faraday waves undergo conical collapse.
Although limited by the temporal resolution of the camera, we witness the same process occurring in our spherical interface scenario.
For the $l=1$ shape mode, the retracting lobe forms an approximately conical shape before ejecting a jet (Extended Data Fig.~\ref{fig:CavityCollapse}a).
Jets arising from a collapsing conical interface are also observed in other problems, including bubble bursting at fluid interfaces \citep{Kientzler1954PhotographicSurface,Duchemin2002JetSurface}, droplet impact on liquid pools \citep{Worthington1897V.Photography,Thoroddsen2018SingularCraters}, cavitation bubble collapse in extremely close proximity to solid boundaries \citep{Lechner2019FastStudy,Reuter2021SupersonicBubbles} and coalescence of bubbles \cite{Jiang2024AbyssCollisions}.
The $l = 2$ shape mode exhibits a more peculiar behaviour due to its axisymmetry (Extended Data Fig.~\ref{fig:CavityCollapse}b).
When the bubble shifts from a prolate to an oblate form, the two retracting lobes at opposite poles become conical—similar to the $l = 1$ case—before emitting two jets that converge at the center.
Conversely, during the transition from oblate to prolate, the equatorial belt retracts annularly, adopting a parabolic profile.
As it ruptures, two jets are ejected, diverging from the center.
The latter behaviour parallels Worthington jets observed when a solid impacts a liquid surface, generating an axisymmetric cavity that pinches off, producing jets directed upward and downward \cite{Gekle2009High-speedImpact,Gekle2010GenerationFormation}.
Both types of collapsing interfaces can be unified under a common theoretical framework for axisymmetric cavity collapse driven by a radial velocity field \cite{Gordillo2023TheoryJets}.
For shape modes with $l>2$, the bubble produces multiple jets from its retracting lobes, each taking on a conical form (Extended Data Fig.~\ref{fig:CavityCollapse}c).
Annular collapses with parabolic profiles are precluded, as modes with $l>2$ are not axisymmetric.

Shape modes typically align one of their symmetry axes with the ultrasound direction (82\% of 67 cases with ultrasound at a 90° angle to the substrate and 80\% of 80 cases at a 45° angle).
This ensures that at least one jet per shape mode period aligns with the ultrasound.
For $l = 1$ and $l = 2$ modes, this results in the jet striking the substrate when the ultrasound is directed towards it.
Although theoretically, modes with $l \ge 3$ should not produce jets reaching the substrate as they converge at the bubble center, deviations from the ideal shape often result in jets hitting the substrate.
We conclude that cyclic jets directed against the substrate occur even without a rigid backing substrate.
Consequently, these jets do not rely on a pressure gradient driver to form but rather on interface instabilities, setting them apart from classical inertial jets \cite{Prentice2005MembraneCavitation}.
Moreover, shape mode-induced jets occur concurrently and repeatedly, unlike inertial jets, which are solitary and transient (Extended Data Fig.~\ref{fig:InertialJetting}, Supplementary Video 4).
They also require approximately ten times lower acoustic pressures to initiate, as shape modes concentrate kinetic energy at single points on the bubble interface where the lobes fold inward.
The formation of shape mode-induced jets remains consistent on the cellular substrate as well (Extended Data Fig.~\ref{fig:Shapemode_Cells}a-b, Supplementary Videos 5 and 6).
Shape modes with  $l = 1$ and $l = 2$ are clearly identified, while higher modes are absent due to the radius of the bubbles studied being smaller than $\SI{5}{\micro\meter}$.
The first two modes are expected to be the most relevant for practical applications, as they manifest for bubble sizes at the sides of the resonant dimension, which, if targeted, allows to minimise the ultrasound pressure used (Fig.~\ref{fig:Fig3}g).
We note that the presence of a significant pressure gradient, such as that caused by a neighbouring rigid substrate, can facilitate the formation of shape mode-induced jets pointing away from the gradient while restraining those directed toward it (Fig.~\ref{fig:Fig1}d).

\newpage

\newpage

\subsection*{Inception threshold for jetting and sonoporation }

Having elucidated the mechanism of jet formation, we now examine how bubble motion and jetting impact the cellular substrate and induce sonoporation.
In our tests ($n = 38$), we observe a critical threshold for the maximum radial expansion of the bubble around \SI{1}{\micro\meter}, independent of the bubble equilibrium radius in the size range examined, beyond which cells experience sonoporation.
In every instance where sonoporation occurred ($n = 19$), the bubble produced cyclic microjets (Fig.~\ref{fig:Fig4}a).
The first microjet within an ultrasound pulse consistently appears when the bubble radial expansion reaches approximately \SI{1}{\micro\meter} (Fig.~\ref{fig:Fig4}b).
These results suggest that sonoporation is enabled by microjetting.
The ultrasound pressure required to exceed the radial expansion threshold for jetting depends on the bubble size and the driving frequency.
For a driving frequency of \SI{1}{\mega\hertz}, the necessary ultrasound pressure amplitude ranges from approximately \SI{50}{\kilo\pascal} for resonant-sized bubbles to around \SI{200}{\kilo\pascal} for non-resonant bubbles (Fig.~\ref{fig:Fig4}c).

Prior studies have reported a similar critical threshold in bubble radial expansion for successful sonoporation when employing microsecond-long ultrasound driving \cite{Helfield2016BiophysicalSonoporation,Beekers2020OpeningSonoporation,Beekers2022InternalizationTunnels}.
However, their top-view perspective on the cell monolayer did not reveal potential bubble jetting.
Considering our analogous experimental conditions and outcomes, cyclic jetting could plausibly be the driving mechanism for sonoporation in those studies as well.
The significance of a threshold in radial expansion becomes clear when interpreted in terms of radial acceleration.
Indeed, it is interfacial acceleration that destabilises the surface between two fluids, possibly leading to jet formation.
By approximating the interfacial acceleration as $a \sim (R_{\rm max} - R_0) \omega^2$, where $R_{\rm max}-R_0$ is the maximum bubble expansion and $\omega$ is the angular driving frequency, we identify a threshold in acceleration for jetting at around \SI{40}{\micro\meter\per\micro\second\squared} (Fig.~\ref{fig:Fig4}a-b).

Common beliefs associate bubble jetting solely with ``inertial cavitation", a regime characterised by violent bubble collapse and fragmentation, typically occurring when the bubble expands beyond twice its equilibrium size. 
However, the cyclic jetting observed here, which results from interface instabilities, predominantly occurs within the ``stable cavitation" regime, where bubbles exhibit a relatively gentle response (Fig.~\ref{fig:Fig4}a-b).

\newpage

\subsection*{Jetting stress evaluation} 

\begin{figure}[!ht] 
    \centering
        \includegraphics[width=\columnwidth]{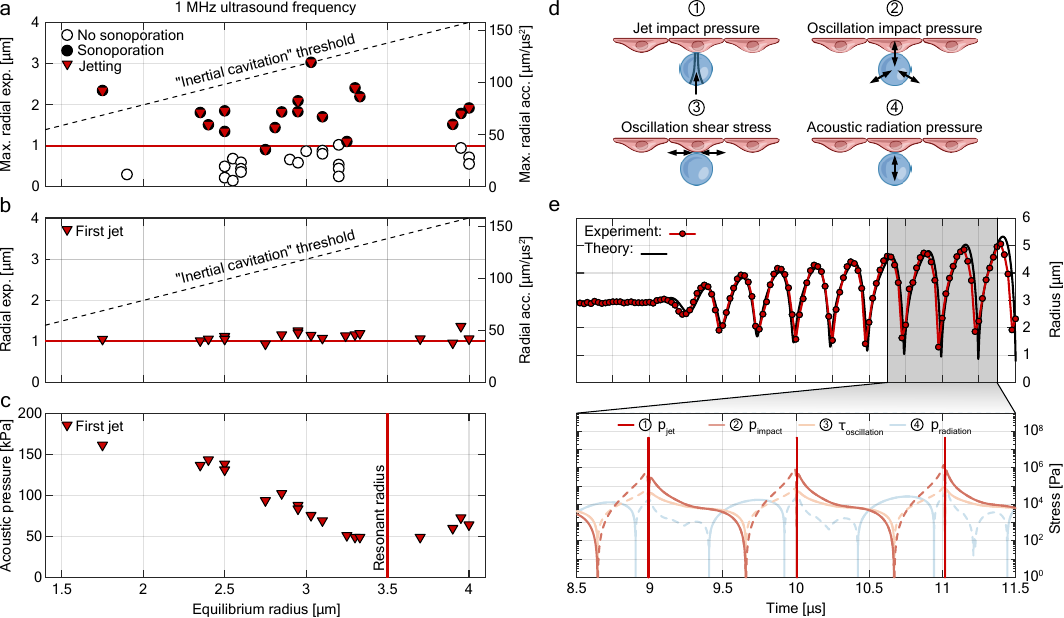}
    \caption{ Cyclic microjetting as a mechanistic threshold indicator for sonoporation. (a) Sonoporation outcome and jetting occurrence as a function of the bubble equilibrium radius, maximum radial expansion and corresponding maximum radial acceleration.
    Cyclic microjetting and sonoporation both occur when the bubble expands beyond approximately \SI{1}{\micro\meter}.
    The threshold for “inertial cavitation”, defined as when a bubble radially expands to twice its equilibrium radius, is also depicted.
    (b) Appearance of the first microjet during the ultrasound pulse as a function of the bubble equilibrium radius, instantaneous radial expansion and corresponding instantaneous radial acceleration.
    The first microjet consistently forms when the bubble expands to at least \SI{1}{\micro\meter}. 
    (c) Appearance of the first microjet during the ultrasound pulse as a function of the bubble equilibrium radius and the instantaneous ultrasound pressure amplitude.
    The minimum pressure required for jetting occurs when the bubble is at its resonant radius. 
    The uncertainty in the bubble radius measurement corresponds to half the pixel size (\SI{80}{\nano\meter}).
    (d) Different stress mechanisms elicited by an ultrasound-driven microbubble on a cell.
    (e) Stress evolution over time in logarithmic scale of each stress mechanism (bottom), inferred from the radial dynamics of the microbubble depicted in Fig.~\ref{fig:Fig1}d and characterised in Fig.~\ref{fig:Fig2} (top).
    Solid lines represent stresses directed towards the substrate or outward from the bubble, whereas dashed lines indicate the opposite.   }
    \label{fig:Fig4}
\end{figure}

Besides microjetting, the bubble can cause mechanical damage to the neighbouring cell layer through several other mechanisms.
These include repeated collisions caused by the bubble oscillation, alternating viscous stresses from the oscillatory flow field, and steady viscous stresses arising from acoustic streaming and penetrating stress exerted by the bubble propelled by Bjerknes forces \cite{Shakya2024Ultrasound-responsiveDelivery} (Fig.~\ref{fig:Fig4}d).

To understand why microjetting is the mechanism enabling sonoporation among all others, we quantify and compare the stress induced by each mechanism.
Upon impact, a fluid jet induces a localised and transient high-pressure region on the cell.
For a jet traveling at speed $U_{\rm jet}$ impacting a substrate with density $ \rho_{\rm l}$ and speed of sound $c_{\rm l}$ comparable to water, the generated pressure can be evaluated using the water hammer pressure formula \cite{Cook1928ErosionWater-hammer,deHaller1933UntersuchungenKorrosionen}, as $p_{\rm jet} \approx  \rho_{\rm l} c_{\rm l} U_{\rm jet} / 2$.
The impact (or suction) pressure exerted on the cell by a bubble expanding (or contracting) at rate $\dot{R}$ can be estimated using the Bernoulli stagnation pressure: $p_{\rm impact}(t) \approx \rho_{\rm l} |\dot{R}| \dot{R} / 2$. 
The tangential viscous stress acting on the cell from the oscillatory flow field can be approximated from the quotient of the bubble wall velocity $\dot{R}$ and the boundary layer thickness $\delta$ \cite{Rooney1972ShearEffects}, as $\tau_{\rm oscillation}(t) \approx {\mu_{\rm l} \dot R}/{\delta}$, where $\delta$ can be determined as $\delta = \left({2 \mu_{\rm l}}/{ \rho_{\rm l} \omega}\right)^{1/2}$, with $\mu_{\rm l}$ being the fluid viscosity.
The oscillatory flow field produces a secondary steady flow known as acoustic streaming. This motion, being a second-order effect, is typically 1-2 orders of magnitude weaker than the primary oscillatory flow field \cite{Longuet-Higgins1998ViscousBubble}.
For coated microbubbles exposed to ultrasound pressures similar to those used in this study, the maximum observed acoustic streaming velocity does not exceed $u_{\rm stream} \approx \SI{0.1}{\meter\per\second}$ \cite{Lajoinie2018Non-sphericalMicrobubbles,Pereno2018LayeredEffects}.
Consequently, the resulting stress, $\tau_{\rm stream} \approx {\mu_{\rm l} u_{\rm stream}}/{\delta}$, remains below a few hundreds of pascals.
Moreover, acoustic streaming initiates only after several cycles of bubble oscillation and can thus be neglected in scenarios involving ultrasound pulses lasting mere tens of microseconds, as in our study.
Finally, the primary and secondary Bjerknes forces generate an averaged pressure on the bubble cross-section \cite{Prosperetti1982BubbleResults,Mettin1997BjerknesField}, expressed as $p_{\rm B1}(t) = -4R\nabla p_{\rm d} / 3$ for the primary Bjerknes force and $p_{\rm B2}(t) = -\rho_{\rm l} R \ddot{V}/12 \pi h^2$ for the secondary Bjerknes force, where $R$ denotes the bubble radius, $\nabla p_{\rm d}$ the gradient of the ultrasound driving pressure, $V$ the bubble volume, and $h$ the distance between the bubble and the plastic substrate.

We assess the stress evolution over time for each damage mechanism (Fig.~\ref{fig:Fig4}e), inferring it from the observed bubble motion in the successful sonoporation test case depicted in Fig.~\ref{fig:Fig1}d and characterised in Fig.~\ref{fig:Fig2}.
The resulting magnitudes of the normal and shear stresses are approximately consistent with previously reported values obtained from three-dimensional finite-element-method  simulations \cite{Bulycheva2024InteractionBoundary,Hosseinkhah2012AMicrovessels}.
The jets produced by the bubble are seen to traverse the entire bubble and impact the substrate within a single frame (\SI{0.1}{\micro\second}), suggesting a jet velocity $U_{\rm jet} > \SI{60}{\meter\per\second}$.
A more accurate measurement is beyond our current experimental capabilities.
At these jet speeds, the resultant jet hammer pressure exceeds that of any other mechanism by at least thirtyfold.
However, this elevated pressure is sustained only briefly, persisting only for the time it takes for a rarefaction wave, generated at the jet contact edge, to reach its central axis.
This duration can be estimated, considering the jet head as spherical \cite{Field2012CavitationThresholds}, as $ \varsigma \sim  R_{\rm jet} U_{\rm jet}/ c^2 > \SI{20}{\pico\second}$, where $R_{\rm jet}$ represents the jet radius.
Subsequently, the pressure along the central axis declines to the lower Bernoulli stagnation pressure, which remains more than fourfold higher than the pressure generated by the side of the bubble impacting the cell.
Furthermore, the pressure exerted by the jet hammer is concentrated over a smaller area compared to other sources of stress \cite{Field2012CavitationThresholds}.
The radius of the contact area can be approximated as $\varrho \sim  R_{\rm jet} U_{\rm jet} / c > \SI{30}{\nano\meter}$.
This value aligns with the pore sizes reported in previous studies that used similar ultrasound driving parameters \cite{Zhou2009TheMembrane,Fan2012SpatiotemporallySonoporation}.
The substantially higher stress generated by the jet upon impact can explain why, for microsecond-long ultrasound driving, only jets enable sonoporation.
However, we cannot rule out the possibility that extended exposure to ultrasound—lasting several milliseconds, as sometimes used in other sonoporation studies \cite{DeCock2015UltrasoundEndocytosis}—may enable weaker stress sources to induce sonoporation.

In conclusion, our study addresses several open questions related to the physics behind drug delivery mediated by microbubbles.
We have elucidated the mechanism of action—cyclic bubble jetting—and the physics behind its formation, modelled bubble and cell dynamics, estimated the mechanical stress generated, and defined thresholds pertinent to bubble dynamics for successful drug delivery.
These findings are expected to significantly advance the development of microbubble-mediated drug delivery systems, accelerating their translation into clinical practice.
Beyond biomedical treatments, the impact of this study extends to diverse technological fields involving bubbles and acoustics, such as sonochemistry, additive manufacturing, and advanced cleaning technologies.

\newpage

\newpage

\section*{Methods}

\subsection*{Cell monolayer culture} 
Primary human umbilical vein endothelial cells (HUVECs, PromoCell GmbH) are cultured in endothelial growth medium (EGM-2, Lonza Bioscience AG) supplemented with 10\% fetal bovine serum (FBS, Gibco\textsuperscript{TM}) within collagen-coated (Corning\textsuperscript{\textregistered} Collagen I, rat tail) T75 flasks at 37°~C in a humidified incubator with 5\% of CO$_2$. 
The medium is changed every three days and cells are passaged before reaching full confluence.
HUVECs used are below passage 12. 
Cells are detached using 0.05\% Trypsin-EDTA (Gibco\textsuperscript{TM}) and seeded on collagen-coated polycarbonate membranes (thickness \SI{50}{\micro\meter}, surface area \SI{1}{\centi\meter\squared}) cut from a CLINIcell (MABIO). 
The membranes are then incubated for four days to form confluent cell monolayers, with the medium being replaced on the third day. 
An example of the resulting cell monolayer is shown in Extended Data Fig.~\ref{fig:Cellmonolayer}.
For imaging it, the cell monolayer is fixed with 4\% paraformaldehyde (PFA, Artechemis), washed with phosphate buffered saline (PBS 1$\times$, Gibco\textsuperscript{TM}), and incubated for thirty minutes at room temperature in 1\% bovine serum albumin (BSA, AppliChem) and 0.3\% Triton X-100 (Sigma Aldrich) in PBS.
Primary cell junction staining is then performed by incubation with mouse anti-CD31 (BD Bioscience) overnight at 4°~C. 
Consequently, cells are washed with PBS for three hours and incubated with AlexaFluor 488 goat-anti-mouse (abcam) for secondary cell junction staining and 4$'$,6-diamidino-2-phenilindol (DAPI, Sigma Aldrich) for nuclei staining overnight at 4°~C. 
HUVECs are then washed for three hours and imaged with a Leica DMI6000B inverted epifluorescence microscope.

\subsection*{Microbubble preparation}
Lipid-coated microbubbles are prepared in-house by probe sonication. 
The gas core is made of \ce{C4F10} (perfluorobutane, Fluoromed) and the lipid coating consists of 90 mol\% {DPPC} (1,2-distearoyl-sn-glycero-3-phosphocholine, NOF EUROPE) and 10 mol\%  of  {DSPE-PEG2K} (1,2-distearoyl-sn-glycero-3-phosphoethanolamine-N-[methoxy(polyethylene glycol)-2000]), Larodan).
The lipids are first dissolved in chloroform, which is then let to evaporate under vacuum at $\SI{35}{\celsius}$ overnight, resulting in the formation of a dry lipid film.
The film is rehydrated with PBS $1\times$ (Boston BioProducts) to yield a total lipid concentration of $ \SI{2}{\milli\gram\per\milli\litre}$ and mixed using a probe sonicator (SFX550, Branson; $\SI{20} {\kilo\hertz}$, $\SI{550} {\watt}$) at low power (30\%).
Microbubbles are formed by probe-sonicating the surface of the lipid solution at full power for ten seconds while simultaneously flowing  \ce{C4F10} gas over it.
The microbubble suspension is then cooled down to room temperature and washed using centrifugation.
Finally, differential centrifugation is employed to isolate microbubbles within a targeted size range (\SIrange{1}{4}{\micro\meter}-radius) \cite{Feshitan2009}.
This size selection has been validated using a particle sizer (Multisizer 4e, Beckman Coulter), which is also used to measure the bubble concentration.

\subsection*{Test chamber preparation} 
A \SI[product-units = single]{10 x 10 x 10}{\milli\metre} water-tight test chamber is used to enclose the cell monolayer (Fig.\ \ref{fig:Fig1}b).
The chamber features transparent acrylic side walls and is open at both the top and bottom.
A silicon rubber cap seals one aperture.
The plastic substrate cultivated with cells is placed in the chamber in contact with the rubber cap.
The chamber is filled with a solution containing PBS 1$\times$, propidium iodide (PI, Sigma-Aldrich, \SI{25}{\micro\gram\per\milli\liter}) and microbubbles ($\approx 500$ microbubbles/mL).
Tape seals the second opening while still allowing the transmission of the ultrasound pulse.

\subsection*{PEG substrate preparation}
PEG hydrogels are prepared following previously described protocols \cite{Dudaryeva20213DDeath,Emiroglu2022BuildingDeformations}.
Briefly, 4-arm poly(ethylene glycol)-norbornene (PEG-NB) polymer precursor solution is mixed with matrix metalloproteinase (MMP)-cleavable peptide linker (KCGPQGIWGQCK, Genscript) and arginylglycylaspartic acid cell adhesion peptide (CRGDS, Genscript) for a final hydrogel solution of 3.25 wt\% PEG-NB, \SI{2.35}{\milli\mole} KCGPQGIWGQCK, and \SI{1.8}{\milli\mole} CRGDS.
To adjust for pH, a small volume of NaOH is added.
This solution is then mixed with PBS and lithiumphenyl-2,4,6-tri-methylbenzoyl-phosphinate (LAP) photoinitiator to reach a final volume of \SI{100}{\micro\liter}.
For the measurement of the compressive elastic modulus of PEG hydrogels via atomic force microscopy (AFM), \SI{10}{\micro\liter}-gel samples are formed between two hydrophobic glass slides, separated by a \SI{0.5}{\milli\meter}-rubber spacer to ensure a flat topography.
Crosslinking is achieved through photoinitiated thiolene chemistry between the thiol groups of the MMP-cleavable linkers and the norbornene groups of PEG–NB.
This is carried out under blue light exposure ($\lambda$ = \SI{405}{\nano\meter}, I = \SI{14.5}{\milli\watt\per\centi\meter\squared}, t = \SI{90}{\second}). 
Gel samples are then let to swell for twenty four hours in Dulbecco's modified Eagle's medium (DMEM) (Gibco, 1230-032) prior to AFM measurements. 
For the experiments with microbubbles, the hydrogel precursor solution is placed on one of the external surfaces of a square glass capillary with a cross section of \SI[product-units = single]{0.90 x 0.90}{\milli\metre} (CM Scientific), which has been pre-treated with plasma.
The capillary serves as a support for the hydrogel substrate.
To maintain the solution atop the capillary during gelation, two rubber delimiters are used.
The resulting thickness of the hydrogel substrate is \SI{2}{\milli\meter}.
Crosslinking is achieved once again through blue light exposure.
The capillary-hydrogel samples are then stored in PBS until use.

\subsection*{Experimental set-up} 
The cell monolayer test chamber, or alternatively, the capillary holding the PEG substrate, is immersed in a water bath filled with deionised water ($T_{\rm l} \approx \SI{22}{\celsius} $).
When employing the PEG substrate, microbubbles are introduced using a syringe beneath the substrate, where they adhere via flotation.
To mitigate potential interference, the microbubbles under examination are situated at least \SI{50}{\micro\meter} away from the test chamber walls or the periphery of the PEG substrate.
For gaining comprehensive insights into the behaviour of single microbubbles interacting with cells and to evaluate the resulting sonoporation outcome, we developed a horizontal microscope (Extended Data Fig.\ \ref{fig:Setup}) which enables ultra-high-speed bright-field and fluorescence recordings of the monolayer with a side-view perspective.
The system is realised using modular optomechanics components (Thorlabs, cage system) and  installed on an optical table  (T1220C, Thorlabs) with active isolators (PTS603, Thorlabs) to minimise environmental vibrations. 
The microscope features a water-dipping objective lens (CFI Plan 100XC W, Nikon) with a focal length of \SI{2}{\milli\meter} and a tube lens (TL400-A, Thorlabs) with a focal length of \SI{400}{\milli\meter}, resulting in a total magnification of $200\times$. 
The terminal section of the microscope, housing the objective lens, is inserted in the water bath through a sealed opening in the water container.
An ultra-high-speed camera (HPV-X2, Shimadzu) allows for recordings at 10 million frames per second over $\SI{25.6}{\micro\second}$ of continuous visualisation of a $\SI{64 x 40} {\micro\meter}$ field of view with a 160-nm pixel resolution.
Backlight illumination is provided by a continuous halogen illuminator (OSL2, Thorlabs) and two Xenon flash lamps operated sequentially (MVS-7010, EG\&G), dedicated for live imaging and video recording, respectively. 
The light sources are all combined into a single optical fiber output and focused on the sample with a custom-built condenser (L1 and L2, AC127-025-A, Thorlabs).
The test chamber or the capillary position is controlled by a three-axis motorised microtranslation stage (PT3/M-Z8, Thorlabs).

A high-intensity focused ultrasound (HIFU) transducer (PA1280, Precision Acoustics; $\SI{1} {\mega\hertz}$ of center frequency, $\SI{75} {\milli\meter}$ of focal length, $\approx \SI{3} {\milli\meter}$ of beam width at $\SI{-6} {\deci\bel}$) is used to drive the microbubbles.
When the test chamber is employed, the transducer is positioned in the water bath at an angle of $75 \degree$ with respect to the horizontal plane to minimise acoustic reflections within the test chamber.
Conversely, when the capillary is employed, the transducer is positioned at either a $90 \degree$ or $45 \degree$ angle.
The driving pulse is generated with a function generator (LW 420B, Teledyne LeCroy) and amplified by a radiofrequency power amplifier (1020L, E\&I). 
A calibrated needle hydrophone ($\SI{0.2} {\milli\meter}$, NH0200, Precision Acoustics) is employed to align the ultrasound focal point with the optical field and to record the shape of the ultrasound pulse envelope, which is utilised as input for the bubble radial dynamics model.
The transducer is manoeuvred using a manual three-axis microtranslation stage (three DTS50/M, Thorlabs).

Fluorescence microscopy is conducted by employing a 532-nm continuous-wave laser (Verdi G10, Coherent) as the excitation light source.
An acousto-optic tunable filter (AOTF.NC-VIS/TN, AA Opto Electronic) acts as an electronic laser shutter.
The transmitted beam ($0^{\rm th}$ order) is ceased while the diffracted beam ($1^{\rm st}$ order) can be activated or deactivated by adjusting the radiofrequency drive power supplied by the AOTF driver (MOD.8C.10.b.VIS, AA Opto Electronic).
To achieve a laser spot size matching the entire field of view, the beam is enlarged by employing a 10$\times$ beam expander (52-71-10X-532/1064, Special Optics).
The laser beam is then focused on the back focal plane of the objective lens by adjusting the spacing of a lens relay system (L3 and L4, AC254-100-A, Thorlabs) in order to achieve a collimated beam emerging from the objective lens.
Undesirable light wavelengths are blocked from reaching the specimen by means of a narrow passband excitation filter (ZET532/10x, Chroma).
The laser beam is steered into the objective lens using a reflective-band dichroic beamsplitter (ZT532dcrb, Chroma).
The laser line is removed from the specimen image with an emission filter (ET590/50m, Chroma).

The activation of the ultrasound pulse, camera recording, light flash, and laser switching are synchronised through a delay generator (DG645, Stanford Research Systems).

\subsection*{Atomic force microscopy measurements of PEG substrates}
PEG hydrogels meant for AFM-based mechanical characterisation are placed on positively-charged glass slides (Superfrost, Thermo Fischer Scientific) for improved adhesion and secured attachment during measurements.
A hydrophobic marker pen is used to trace the area surrounding the hydrogel and \SIrange{300}{400}{\micro\liter} of PBS at room temperature is used to fully immerse the gel, ensuring hydration during the indentation measurements.
Nanoindentation measurements are performed using a Flex-Bio AFM (Nanosurf, Switzerland).
The hydrogels surface is indented using a colloidal probe made up of a cantilever with a nominal spring constant $k = \SI{0.1}{\newton\per\meter}$ and a \SI{10}{\micro\meter}-diameter borosilicate glass bead affixed to the cantilever tip (CP-qp-CONT-BSG-B-5; Nanosensors).
Prior to testing, the spring constant is calculated using the Sader method, as implemented in the standard Nanosurf software.
The slope of the deflection-displacement curve, obtained from the indentation of a bare region of the glass slide, is used to determine the deflection sensitivity.
The AFM is mounted on top of an inverted microscope (Nikon Eclipse Ti-E) to allow for sample visualisation and macroscopic positioning of the probe.
The deflection and displacement of the cantilever are recorded as the probe descended, indented the gel surface at a speed of \SI{1}{\micro\meter\per\second}, and then retracted, producing a force-displacement curve for each indentation.
Between four and five random locations, each measuring \SI[product-units = single]{50 x 50}{\micro\metre}, are selected for indentation on each sample.
At each location, measurements are taken at a grid of 5 $\times$ 5 points.
To obtain the apparent compression elastic modulus $E$ from each force-displacement curve, and in accordance with current standard practices in AFM nanoindentation, we employ the Hertz contact model, which is considered the most appropriate for a sphere indenting a semi-infinite half-space:
\begin{equation}
F = \frac{4}{3}\frac{E}{1-\vartheta^2}\mathcal{R}^{1/2}\iota^{3/2},
\end{equation}
where $F$ is the force applied by the cantilever, $\vartheta$ is the Poisson’s ratio, $\mathcal{R}$ is the radius of the spherical bead, and $\iota$ is the indentation. 
For consistency with the literature and simplification, the Poisson’s ratio is considered to be $0.5$.
Each curve is fitted using a custom-built python-based algorithm to extract the apparent elastic modulus.
Force-displacement curves without a clear contact point are discarded.
The mean apparent compression elastic modulus of the PEG hydrogels is $E = \SI[multi-part-units = single]{0.55(10)}{\kilo\pascal}$.
This value is obtained by averaging the mean values of the median apparent modulus values in 4-5 locations in each hydrogel ($n=3$).

\subsection*{Image analysis} 
The time evolution of the microbubble radius and position is extracted from the bright-field recordings using an image analysis script written in MATLAB.
Single frames are first filtered with a median filter to remove image noise and then binarised using locally adaptive thresholding.
Subsequently, a flood-fill operation on holes in the binarised image is performed and all small noisy connected components are removed.
The time-varying contour of the binarised bubble, denoted as $R(z,t)$ with $z$ representing the axis normal to the substrate, is isolated.
Since the bubble may undergo a non-spherical collapse—albeit predominantly axisymmetric—an equivalent bubble radius is derived by integrating the cross-sectional area, as follows:
\begin{equation}
R_{\rm eq}(t) = \left( \frac{3}{4} \int_{z^{-}}^{z^{+}}  R(z,t)^2 dz \right)^{1/3},
\end{equation}
where $z^{-}$ and $z^{+}$ represent the limiting values within which the bubble contour $R(z,t)$ is defined.
Finally, the bubble position is extracted by computing the centroid of binarised bubble image.

\subsection*{Theoretical modelling of bubble radial dynamics} 

The radial motion of a coated microbubble, assumed spherical with a radius $R(t)$, is described using the model introduced in our earlier study \cite{Cattaneo2023ShellMicrobubbles}.

In brief, the dynamics of the liquid around the bubble is modelled using the Rayleigh--Plesset equation for mildly compressible Newtonian media \cite{Brenner2002Single-bubbleSonoluminescence}, which reads:
\begin{equation}\label{eq:RP}
\rho_{\rm l} \left(R \ddot R + \frac{3}{2} \dot R^2 \right)= \left( 1 + \frac{R}{c_{\rm l}} \frac{d}{dt} \right) p_{\rm g}  + \Sigma\left(R, \dot R\right) - p_{\infty} - p_{\rm d}\left(t\right) - 4\mu_{\rm l} \frac{\dot R}{R},
\end{equation}
where over-dots denote time differentiation, $\rho_{\rm l} = \SI{997.8} {\kilo\gram\per\cubic\meter}$ is the liquid density, $c_{\rm l} = \SI{1481} {\meter\per\second}$ is the speed of sound in the medium, $p_{\rm g}$ is the gas pressure inside the bubble, $\Sigma(R, \dot R)$ is the pressure term that accounts for the generalised interfacial stresses, $p_{\infty} = \SI{102.2} {\kilo\pascal}$ is the undisturbed ambient pressure, $p_{\rm d}(t)$ is the ultrasound driving pressure and $\mu_{\rm l}=\SI{9.54e-4} {\pascal\second}$ is the dynamic viscosity of the medium.

The ultrasound pressure $p_{\rm d}(t)$ can be expressed as the product of the signal amplitude $p_{\rm a}$ and the normalised time signal $\phi(t)$:
\begin{equation}
p_{\rm d}(t) = p_{\rm a}\phi(t).
\end{equation}
$\phi(t)$ is  measured experimentally using a hydrophone positioned at the ultrasound focal point without the presence of the test chamber, while $p_{\rm a}$ serves as the sole fitting parameter in our theoretical model.
This indirect method of measuring the ultrasound pressure amplitude allows to account for the variable acoustic absorption caused by neighboring bubbles in the test chamber, as well as the acoustic reflections at the bubble’s location.
Direct measurement using a hydrophone would be impractical due to the minute scale of the problem. 
Regarding acoustic reflections, the rigid substrate used to cultivate the cells reflects part of the incoming acoustic wave, which then interferes with the incident wave, potentially with a phase shift.
However, since the acoustic wavelength is over 100 times larger than the distance between the bubble and the substrate, this distance is negligible relative to the wavelength.
As a result, the reflected wave remains in phase with the incident wave at the bubble’s location.
Thus, from the bubble’s perspective, the rigid substrate merely amplifies the pressure of the incident wave, potentially doubling it if the substrate is a perfect reflector.
Therefore, fitting the model based solely on the pressure amplitude $p_{\rm a}$ is sufficient to account for any acoustic reflections at the bubble's location.

The phospholipid coating reduces the gas-liquid surface tension, which decreases the large Laplace pressure at these bubble sizes, thereby halting gas efflux and prolonging the bubble’s lifespan.
Additionally, it imparts rheological properties to the interface \cite{Edwards1991InterfacialRheology}.
The interfacial pressure term $\Sigma(R, \dot R)$ related to the phospholipid coating is described with the Marmottant model \cite{Marmottant2005ARupture}, which is articulated as follows:
\begin{multline}
 \Sigma\bigl(R, \dot R\bigr) = - 2\frac{\sigma (R)}{R} -4\kappa_{\rm s}\frac{\dot R}{R^2}, \\
\label{eq:InterfacePressureMarmottant}
     \text{with} \ \  \sigma\bigl(R) = 
\begin{cases}
 0,  \! \! \! \! & \text{for } R\leq R_{\rm buckling},\\
   {\sigma_0 + E_{\rm s} \left(J-1\right)},  \! \! \! \! & \text{for } R_{\rm buckling}< R \leq R_{\rm rupture},\\
    {\sigma_{\rm water}},  \! \! \! \! & \text{for } R > R_{\rm rupture},
\end{cases}
\end{multline}
where $\sigma(R)$ is the interfacial surface tension and $\kappa_{\rm s}$ is the interfacial dilatational viscosity. 
$\sigma(R)$ is a piecewise-defined function that accounts for: (i) the buckling of the coating, which occurs when the bubble is compressed to a radius smaller than $R_{\rm buckling}$, resulting in a tension-free interface; (ii) the rupture of the coating, which occurs as the bubble expands beyond $R_{\rm rupture}$, leading to the complete exposure of the gas core to the surrounding fluid, thus establishing a surface tension equivalent to that of a clean gas-water interface ($\sigma_{\rm water} = \SI{72.8}{\milli\newton\per\meter}$); and (iii) the elastic regime characterised by an interfacial dilatational modulus $E_{\rm s}$ that lies in between. $\sigma_0$ is the interfacial surface tension at equilibrium and  $J=R^2 / R_0^2$ is the relative area deformation.
The values for the shell rheological parameters are specified based on our prior rheological investigation \cite{Cattaneo2023ShellMicrobubbles} and read $\kappa_{\rm s} = \SI{5e-9} {\kilo\gram\per\second}$, $E_s = \SI{0.2} {\newton\per\meter}$ and $\sigma_0 = \SI{0} {\newton\per\meter}$.
A tensionless bubble at equilibrium is consistent with its observed long-term stability against dissolution in a saturated medium \cite{Ferrara2007UltrasoundDelivery}.

For microbubbles driven by ultrasound at megahertz frequencies, the validity of the commonly-used polytropic process approximation for the bubble gas pressure $p_{\rm g}$ is typically violated because the Péclet number is close to one \cite{Prosperetti1988NonlinearDynamics}.
Therefore, we employ Zhou's model \cite{Zhou2021ModelingBubble} to address the thermal interaction problem which governs the bubble gas pressure $p_{\rm g}$.
The method is based on the well-accepted assumption that the gas pressure is uniform within the bubble  \cite{Prosperetti1988NonlinearDynamics}.
This allows, for a perfect gas, to exactly express the gas radial velocity as:
\begin{equation}\label{eq:ug}
u_{\rm g}(r) = \frac{1}{\gamma p_{\rm g}}\left[\left(\gamma-1\right) K_{\rm g} \frac{\partial T_{\rm g}}{\partial r} -\frac{1}{3} r \dot p_g \right],
\end{equation}
and from this result recover an exact relation for the gas pressure:
\begin{equation}\label{eq:pg}
\dot p_{\rm g} = \frac{3}{R}\left[\left(\gamma-1\right) K_{\rm g} \left.\frac{\partial T_{\rm g}} {\partial r}\right|_R -\gamma p_{\rm g} \dot R \right],
\end{equation}
where $r$ is the radial coordinate, $\gamma$ is the gas specific heat ratio, $K_{\rm g}$ is the gas thermal conductivity and $T_{\rm g}$ is the gas temperature.
The temperature profile $T_{\rm g}(r)$ inside the bubble is divided into three regions: (1) an internal layer characterised by uniform temperature, (2) a buffer layer, and (3) an external layer characterised by a linear temperature distribution.
The change of bubble surface temperature is significantly smaller compared to that of the gas and can, therefore, be neglected, i.e. $T_{\rm g}|_R \approx T_{\rm l}$ \cite{Prosperetti1988NonlinearDynamics}.
The volume-averaged temperature $T_{\rm g_{\textit{i}}}$ of each region $i$ can be computed through the equation of state for an ideal gas:
\begin{equation}\label{eq:idealgaslaw}
 T_{\rm g_{\textit{i}}} = \frac{p_{\rm g}}{\rho_{\rm g_{\textit{i}}}\mathcal{R}}, \quad \text{for } i = 1,2,3, 
\end{equation}
where $\rho_{\rm g_{\textit{i}}}$ is the volume-averaged gas density of the region $i$ and $\mathcal{R}$ is the specific gas constant.
$\rho_{\rm g_{\textit{i}}}$ can be computed using the continuity equation for each region:
\begin{equation}\label{eq:discretecontinuity}
\dot m_{\rm g,1} = -f_1, \ \ \ \dot m_{\rm g,2} = f_1 - f_2, \ \ \ \dot m_{\rm g,3} = f_2,
\end{equation}
where $m_{\rm g_{\textit{i}}}$ is the gas mass in region $i$ and $f_{j}$ is the mass flux across the region interface $j$, which has the form:
\begin{equation}
 f_j = \rho_{\rm g,\textit{j}}^{\rm uw} u_{\rm g,\textit{j}}^{\rm rel} S_j,  \quad \text{for } j = 1,2,
\end{equation}
where $\rho_{\rm g,\textit{j}}^{\rm uw}$ is the density of the neighbouring cell on the upwind side of interface $j$, $u_{\rm g,\textit{j}}^{\rm rel}$ is the convective velocity, which is the difference between the real velocity of the gas, as defined in Eq.~\eqref{eq:ug}, and the velocity of the cell interface $j$, and $S_j$ is the surface area of the interface $j$.
Due to the long residence time (approximately ten minutes) in an air-saturated environment, the microbubble gas core is replaced by air \cite{Kwan2012LipidExchange}.
Therefore, the values for the gas parameters are $\gamma = 1.4$, $K_{\rm g} = \SI{0.026} {\watt\per\meter\per\kelvin}$ and $\mathcal{R} = \SI{287} {\joule\per\kilo\gram\per\kelvin}$.

\subsection*{Theoretical modelling of bubble displacement dynamics} 

A microbubble, assumed spherical with a radius $R(t)$, in contact with a soft substrate and driven by an ultrasound pulse directed normal to the substrate, displays a translational motion $x(t)$, assumed positive if towards the substrate, that can be described with the force balance equation:

\begin{align}\label{eq:transmotion}
\underbrace{F_{\rm I}(t)}_\text{Inertial force} = \quad \  \underbrace{F_{\rm B1}(t) + F_{\rm B2}(t)}_\text{Acoustic driving forces} \quad \ +  \ \  \quad \underbrace{F_{\rm AM}(t) + F_{\rm VD}(t)}_\text{Hydrodynamic forces} \quad \ \ + \underbrace{F_{\rm CR}(t).}_\text{Cell resistive force}
\end{align}

\noindent $F_{\rm I}(t)$ is the inertial force of the bubble, characterised by its mass $m$:
\begin{equation}
F_{\rm I}(t) = m\ddot{x}.
\end{equation}

\noindent $F_{\rm B1}(t)$ is the primary Bjerknes force induced by the ultrasound driving pressure gradient  $\nabla p_{\rm d}$ on a bubble of volume $V$ \cite{Prosperetti1982BubbleResults}:
\begin{equation}
F_{\rm B1}(t) = -V\nabla p_{\rm d}.
\end{equation}

\noindent $F_{\rm B2}(t)$ is the secondary Bjerknes force caused by the presence of a rigid plastic substrate at a distance $L-x$, where $L$ is the initial thickness of the cell \cite{Mettin1997BjerknesField}.
In accordance with potential flow theory, this plastic substrate can be represented by a virtual bubble that mirrors the real one and, therefore, emits a sound field:
\begin{equation}
F_{\rm B2}(t) = -\frac{\rho_{\rm l} V}{16 \pi (L-x)^2} \ddot{V}.
\end{equation}
In the example reported in Fig.~\ref{fig:Fig1}, the cell thickness amounts to $L = \SI{12}{\micro\meter}$.

\noindent $F_{\rm AM}(t)$ is the added mass force that accounts for the additional fluid mass that gets carried along by the bubble as it moves through the fluid \cite{Takemura2004TheNumber}:
\begin{equation}
F_{\rm AM}(t) = -\frac{1}{2} \rho_{\rm l} \left(\dot{V}x + V\dot{x} \right).
\end{equation}

\noindent $F_{\rm VD}(t)$ is the quasi-steady viscous drag force experienced by the translating bubble \cite{Takemura2004TheNumber}:
\begin{equation}
F_{\rm VD}(t) = - \frac{1}{2} \rho_{\rm l} \pi R^2 \dot{x}^2 C_{\rm D},
\end{equation}
where the drag coefficient $C_{\rm D}$ is taken as \cite{White1991ViscousFlow}:
\begin{equation}
C_{\rm D} = \frac{24}{\rm Re} + \frac{6}{1+ \sqrt{\rm Re}} +0.4,
\end{equation}
with $\text{Re} = 2 \rho_{\rm l} R \dot{x}/ \mu_{\rm l}$ denoting the instantaneous value of the translational Reynolds number. 
The history contribution of the viscous drag force can be neglected at the time-averaged Reynolds numbers, $\widetilde{\text{Re}}$ and $\widetilde{\mathcal{U}\text{Re}}$, with $\mathcal{U} = \dot{R} / \dot{x}$, encountered in the experiments ($\widetilde{\text{Re}} = 0.9$ and $\widetilde{\mathcal{U}\text{Re}}= 16.5$ for the one reported in Fig.~\ref{fig:Fig1}c and $\widetilde{\text{Re}} = 4.3$ and $\widetilde{\mathcal{U}\text{Re}}= 32.7$ for the one reported in Fig.~\ref{fig:Fig1}d) \cite{Magnaudet1998TheRadius}.

\noindent $F_{\rm CR}(t)$ is the cell resistive force:
\begin{equation}
F_{\rm CR}(t) = - \pi R^2 c_{\beta} \frac{\text{d}^{\beta} \epsilon}{\text{d}t^{\beta}},
\end{equation}
where $\epsilon = x/L$ is the cell strain.
Here, we use a single fractional unit, also called a spring-pot, to characterise the rheological behaviour of the living cell \cite{Bonfanti2020FractionalMaterials}. 
A spring-pot leverages the concept of fractional derivative to capture behaviours that lie between those of a spring (${\beta}=0$) and a dashpot (${\beta}=1$).
Our approach is inspired by recent validated models for living cells that employ two parallel fractional units to describe the different behaviours at low and high deformation rates \cite{Hurst2021IntracellularDivision}.
We simplify this framework by only employing the fractional unit associated with high deformation rates as the dynamics under consideration involve exceptionally rapid  deformation ($10^5-10^7 \ \text{s}^{-1}$).
Based on microrheological studies performed on epithelial cells \cite{Hurst2021IntracellularDivision}, we adopt $\beta = 0.8$ and $c_{\beta} = 1 \ \text{Pa  s}^{\beta}$.
A fractional derivative of order $\beta$ with respect to time can be defined according to the Caputo approach \cite{Caputo1967LinearIndependentII} as:
\begin{equation}
    \frac{\text{d}^{\beta} \epsilon(t)}{\text{d}t^{\beta}} = \frac{1}{\Gamma(n-\beta)}\int_0^t(t-s)^{(n-\beta-1)}\frac{\text{d}^n \epsilon (s)}{\text{d}s^n} \ \text{d}s, \quad n-1<\beta<n,
\end{equation}
where $n$ is an integer number and $\Gamma$ is the Gamma function.
For $0<\beta<1$, a simplified expression can be derived:
\begin{equation}
    \frac{\text{d}^{\beta} \epsilon(t)}{\text{d}t^{\beta}} = \frac{1}{\Gamma(1-\beta)}\int_0^t(t-s)^{(-\beta)}\frac{\text{d}\epsilon (s)}{\text{d}s} \ \text{d}s, \quad 0<\beta<1,
\end{equation}
We use the L1 discretisation \cite{Li2019TheoryDerivatives} to numerically compute the Caputo derivative at discrete time points $t_{j} = j \tau$ with $j = 0, 1, 2, \dots, N$: 
\begin{equation}
\left.\frac{\text{d}^{\beta} \epsilon(t)}{\text{d}t^{\beta}}\right|_{t_{k}} = \frac{\tau^{-\beta}}{\Gamma(2-\beta)}\sum_{j=1}^k
b_{k-j}\left(\epsilon(t_{j})-\epsilon(t_{j-1})\right),
\end{equation}
where $b_{k-j}= (k-j+1)^{1-\beta} - (k-j)^{1-\beta}$.

\noindent The gravity effects are disregarded in the force balance due to their negligible contribution.

\subsection*{Derivation of the dimensionless impulse for ultrasound-driven bubbles} 

A bubble, initially at rest in a aqueous medium, of radius $R_0$ and internal pressure $p_{\rm g} + p_{\rm v}$, where $ p_{\rm v}$ is the vapour pressure, exposed to a smooth pressure field $p(\boldsymbol{x})$, described by the approximation $p(\boldsymbol{x}) = p_0 +  \boldsymbol{\nabla} p  \cdot \boldsymbol{x} + O(\boldsymbol{x}^2)$ in the vicinity of the bubble, develops a jet directed inwards against the local pressure gradient $\boldsymbol{\nabla} p$ when the driving pressure jump $\Delta p = p_0 - p_{\rm g} - p_{\rm v} + 2 \sigma_0 / R_0 > 0$.

Obreschkow \textit{et al.} \cite{Obreschkow2011UniversalBubbles} have shown that, for collapsing laser-induced millimetric vapour bubbles exposed to a uniform gravity-induced pressure gradient, the normalised jet volume correlates with the pressure anisotropy parameter $\boldsymbol{\zeta}$, defined as:
\begin{equation}
\boldsymbol{\zeta} = - \boldsymbol{\nabla} p R_{0} / \Delta p.
\end{equation}
In this context, $\boldsymbol{\nabla} p = \rho_{\rm l} \boldsymbol{g}$, where $\boldsymbol{g}$ is the gravitational acceleration, $R_{0}$ stands for the bubble radius at the onset of collapse, while $\Delta p = p_0 - p_{\rm v}$, as the surface tension effects are neglected due to the inertia-dominated dynamics.
$\boldsymbol{\zeta}$ can be regarded as the dimensionless counterpart of the Kelvin impulse $\boldsymbol{I}$, which measures the linear momentum acquired by the fluid during the expansion and subsequent compression phase of the bubble:
\begin{equation}
\boldsymbol{I} = -\int_{T} \int_{S} p(\boldsymbol{x}) \boldsymbol{\hat{n}} \ dS \ dt \simeq - \int_{T} V \boldsymbol{\nabla} p \ dt,
\end{equation}
where $T$ is the expansion and compression time period, $S$ is the bubble surface, $\boldsymbol{\hat{n}}$ is the outward normal to the fluid.
For a collapsing vapour bubble subjected to a constant pressure gradient (such as that induced by gravity), the Kelvin impulse $\boldsymbol{I}$ reads: 
\begin{equation}\label{eq:Idp}
\boldsymbol{I} \simeq - 4.789 R_0^4 \boldsymbol{\nabla} p \sqrt{\rho / \Delta p },
\end{equation}
and it is therefore linked to $\boldsymbol{\zeta}$ through the relation:
\begin{equation}\label{eq:Izeta}
\boldsymbol{I} \simeq  4.789 R_0^3\sqrt{\rho \Delta p}\boldsymbol{\zeta}.
\end{equation}

Supponen \textit{et al.} \cite{Supponen2016ScalingBubbles} have experimentally found that $\boldsymbol{\zeta}$ governs numerous other dimensionless jet parameters of collapsing vapour bubbles via power laws. 
Furthermore, this holds true irrespective of the type of jet driving, namely gravity, free boundaries, rigid boundaries, or any combination thereof.
However, in the presence of a boundary, the pressure gradient $\boldsymbol{\nabla} p(t)$ varies in time during the growth and collapse of a bubble, making it challenging to define $\boldsymbol{\zeta}$ as an integral quantity akin to the Kelvin impulse $\boldsymbol{I}$.
Therefore, the authors defined an equivalent $\boldsymbol{\zeta}$ for neighbouring boundaries such that Eq.~(\ref{eq:Izeta}) still returns the correct Kelvin impulse.

For ultrasound-driven bubbles, both the pressure gradient $\boldsymbol{\nabla} p(t)$ and the driving pressure jump $\Delta p(t)$ vary in time.
Hence, we cannot directly derive the value of $\boldsymbol{\zeta}$ from Eq.~(\ref{eq:Izeta}) due to the dependence of the prefactor on a time-invariant $\Delta p$.
To proceed, we need to establish an equivalent, time-invariant, $\overline{\Delta p}$ for every ultrasound cycle to align ourselves with the framework in which Eq.~(\ref{eq:Izeta}) was derived.
We can do this by recognising that the driving pressure jump $\Delta p(t)$ is unaffected by the existing pressure gradient $\boldsymbol{\nabla} p(t)$, enabling us to effortlessly determine it from the virtual impulse $\widehat{\boldsymbol{I}}$ generated by a discretionary pressure gradient of constant known value, even unitary, $\widehat{\boldsymbol{\nabla} p}$ by means of Eq.~(\ref{eq:Idp}), as follows:
\begin{equation}
\overline{\Delta p} = ( 4.789 R_0^4 \widehat{\boldsymbol{\nabla} p} \sqrt{\rho} / \widehat{\boldsymbol{I}} )^2,
\end{equation}
where $R_0$ is the largest radius reached by the bubble during the ultrasound cycle, akin to the definition of $R_0$ for a collapsing vapour bubble.
At this point, $\boldsymbol{\zeta}$ can be promptly calculated employing Supponen \textit{et al.}'s approach, by deriving it from the actual impulse acquired by the fluid within the ultrasound cycle $\boldsymbol{I}$ by means of Eq.~(\ref{eq:Izeta}), as follows:
\begin{equation}
\boldsymbol{\zeta} \simeq \boldsymbol{I} / 4.789 R_0^3\sqrt{\rho \overline{\Delta p}}.
\end{equation}

\section*{Data availability}
Source data are provided with this paper. These data are also available
via Zenodo at \url{https://doi.org/10.5281/zenodo.14262735}.

\section*{Code availability}
The codes supporting this study are available via GitHub at \url{https://github.com/cttnmrc/jetting-enables-sonoporation.git}.

\section*{Acknowledgments}
We acknowledge funding from the Swiss National Science Foundation (project no. 200021\_200567) and from ETH Zürich (research grant no. 1-010206-000).
L.G.P. acknowledges funding from ETH Zürich (Open ETH project SKINTEGRITY.CH) and the Swiss National Science Foundation (Sinergia project CRSII5\_213498). 
M.L.N. acknowledges funding from the Swiss National Science Foundation (project no. 205321\_188828). 
The funders had no role in study design, data collection and analysis, decision to publish or preparation of the manuscript.

\section*{Author contributions}
M.C. and O.S. conceived the study. M.C., G.S., G.G., L.A.K. and O.S.
defined the methodology. M.C. designed the experimental setup.
G.G. and L.A.K. cultured the cells. G.S. and G.G. prepared the
microbubbles. L.G.P. and G.G. fabricated the PEG substrates. M.L.N.
performed the AFM measurements. M.C., G.G. and G.S. performed the
experiments. M.C. and O.S. analysed and interpreted the data. M.C.
carried out the theoretical modelling and numerical simulations. M.C.
wrote the initial draft of the manuscript. O.S. supervised the research.
All authors contributed to the critical review and revision of the
manuscript.

\section*{Competing interests}
The authors declare no competing interests.

\section*{Additional information}

\noindent\textbf{Correspondence and requests for materials} should be addressed to M.C. and O.S.

\newpage

\section*{Extended Data}
\setcounter{figure}{0}
\renewcommand\figurename{Extended Data Fig.}

\begin{figure}[!ht] 
    \centering
        \includegraphics[width=\columnwidth]{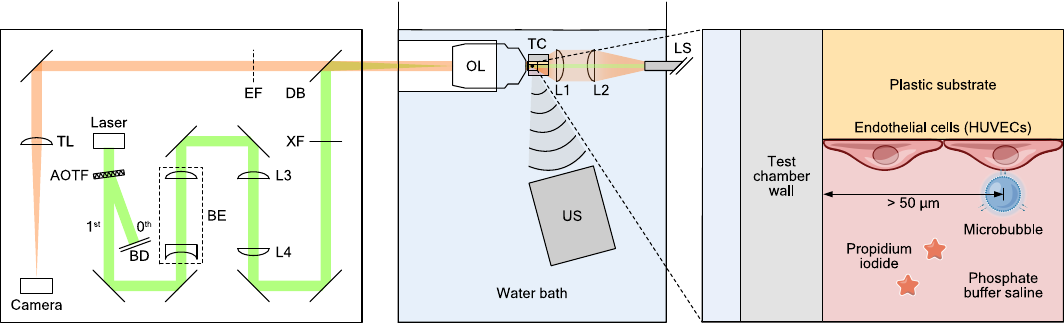}
    \caption{Schematic of the experimental setup. (AOTF) Acousto-optic tunable filter, (BD) Beam dump, (BE) Beam expander, (DB) Dichroic beamsplitter, (EF) Emission filter, (L1-L4) Lens, (LS) Light source, (OL) Objective lens, (TC) Test chamber, (TL) Tube lens, (US) Ultrasound transducer, (XF) Excitation filter.}
    \label{fig:Setup}
\end{figure}

\begin{figure}[!ht] 
    \centering
        \includegraphics[width=\columnwidth]{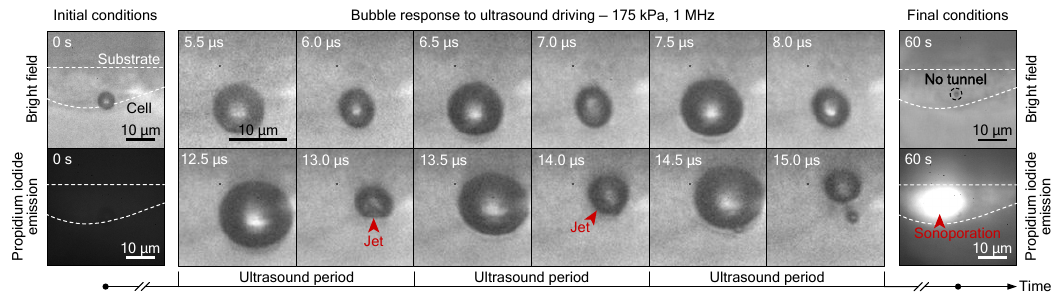}
    \caption{Cyclic microjets generated by a microbubble ($R_0=\SI{3.0}{\micro\meter}$) induce cell membrane poration, facilitating drug uptake ($p_{\rm a} = \SI{175}{\kilo\pascal}, \ f = \SI{1}{\mega\hertz}$).
    Yet, in this instance, the bubble motion does not form a transendothelial tunnel, as the cell deformation recovers and the bubble returns to its initial position when the ultrasound pulse stops.  
    }
    \label{fig:FigNoTunnel}
\end{figure}

\begin{figure}[!ht] 
    \centering
        \includegraphics[width=\columnwidth]{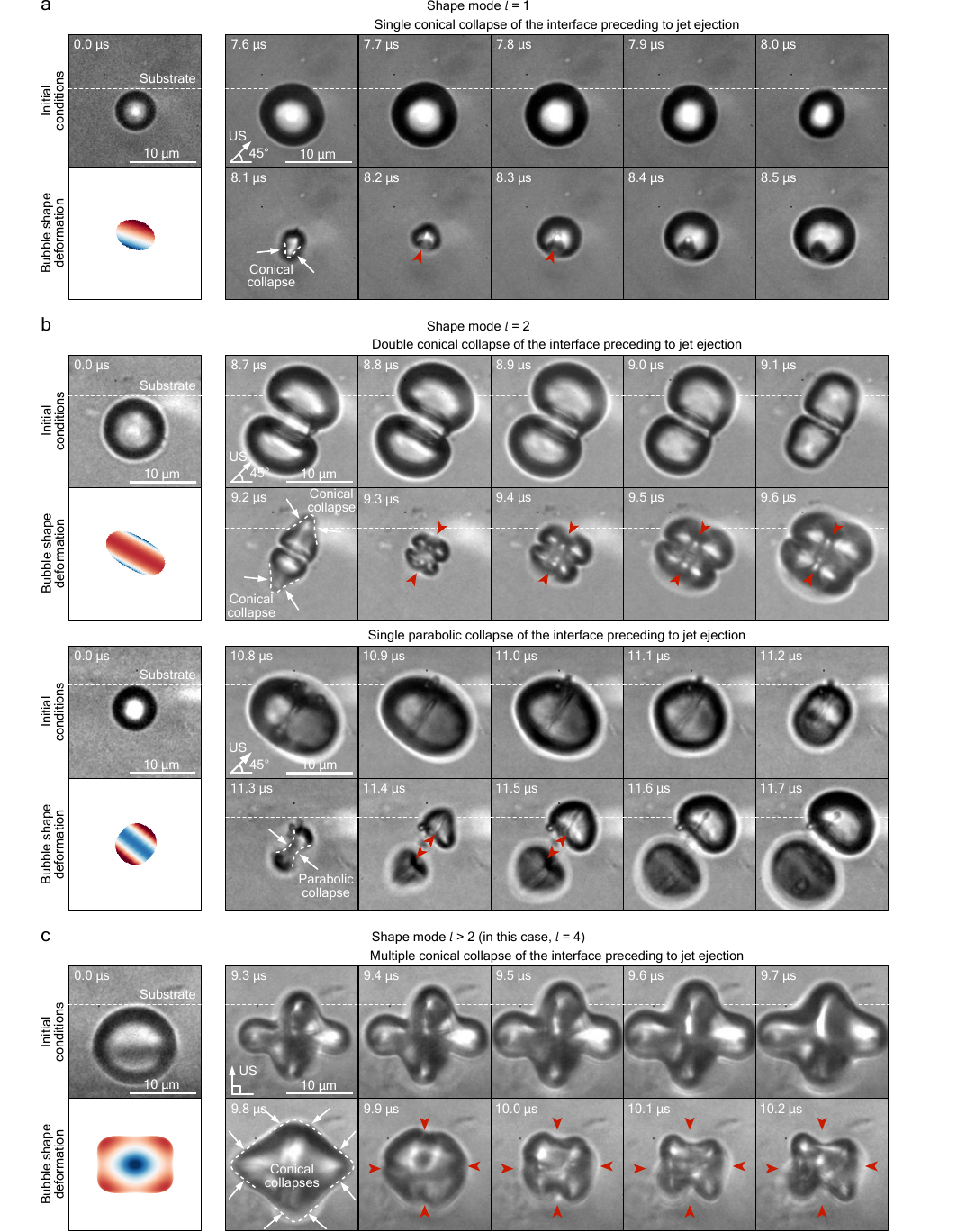}
    \caption{Shape of the bubble interface before jet emission.
    The red arrows indicate the jetting occurrence.
    (a) For the $l = 1$ shape mode, the single bubble lobe collapses taking a conical shape before ejecting a jet.
    (b) For the $l = 2$ shape mode, as the bubble transitions from a prolate to an oblate form, the two bubble lobes at opposite poles follow a conical collapse, releasing two converging jets.    
    Conversely, when the bubble shifts from oblate to prolate, the equatorial region collapses annularly, adopting a parabolic profile.
    Two diverging jets are shoot up- and downwards from the closure point.
    (c) For $l>2$ shape modes ($l=4$ in the present case), the multiple bubble lobes undergo conical collapses, each producing a jet.
    }
    \label{fig:CavityCollapse}
\end{figure}

\begin{figure}[!ht] 
    \centering
        \includegraphics[width=\columnwidth]{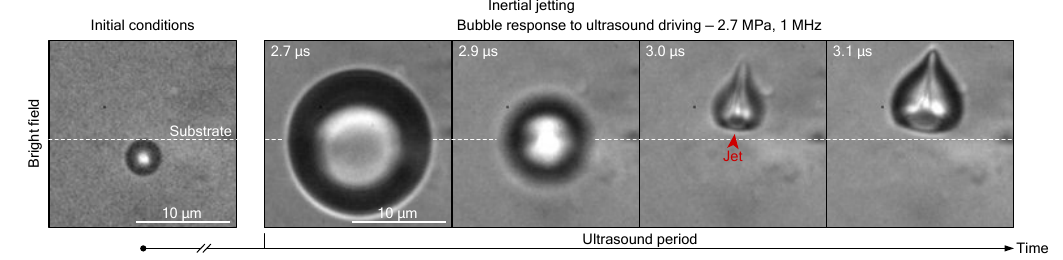}
    \caption{Inertial microjetting of a microbubble ($R_0=\SI{2.3}{\micro\meter}$) in contact with a PEG substrate driven by a very intense ultrasound pulse ($p_{\rm a} = \SI{2.7}{\mega\pascal}, \ f = \SI{1}{\mega\hertz}$). 
    The jet is transient as it is followed by bubble fragmentation.
    }
    \label{fig:InertialJetting}
\end{figure}

\begin{figure}[!ht] 
    \centering
        \includegraphics[width=\columnwidth]{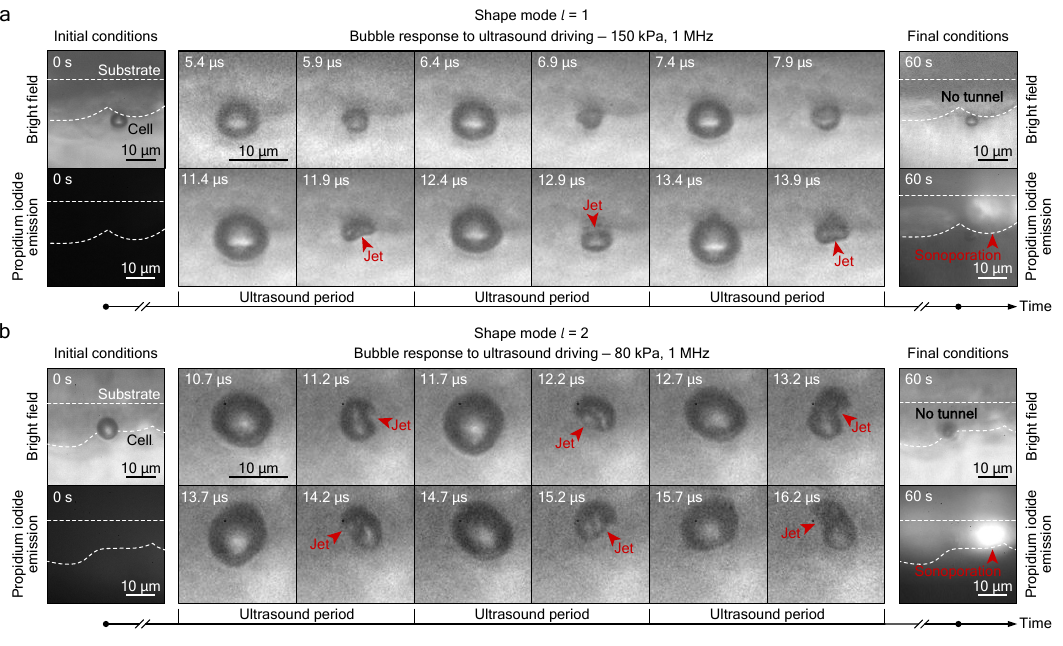}
    \caption{Cyclic microjetting on the cellular substrate resulting in cell membrane poration and drug uptake is driven by shape modes. 
    (a) Jetting induced by a shape mode with wavenumber $l=1$, as the bubble ($R_0=\SI{2.9}{\micro\meter}$) displays an alternate body motion ($p_{\rm a} = \SI{150}{\kilo\pascal}, \ f = \SI{1}{\mega\hertz}$).
    (b) Jetting induced by a shape mode with wavenumber $l=2$, as the bubble ($R_0=\SI{3.8}{\micro\meter}$) alternately takes on a prolate and oblate shape ($p_{\rm a} = \SI{80}{\kilo\pascal}, \ f = \SI{1}{\mega\hertz}$).
    }
    \label{fig:Shapemode_Cells}
\end{figure}

\begin{figure}[!ht] 
    \centering
        \includegraphics[width=\columnwidth]{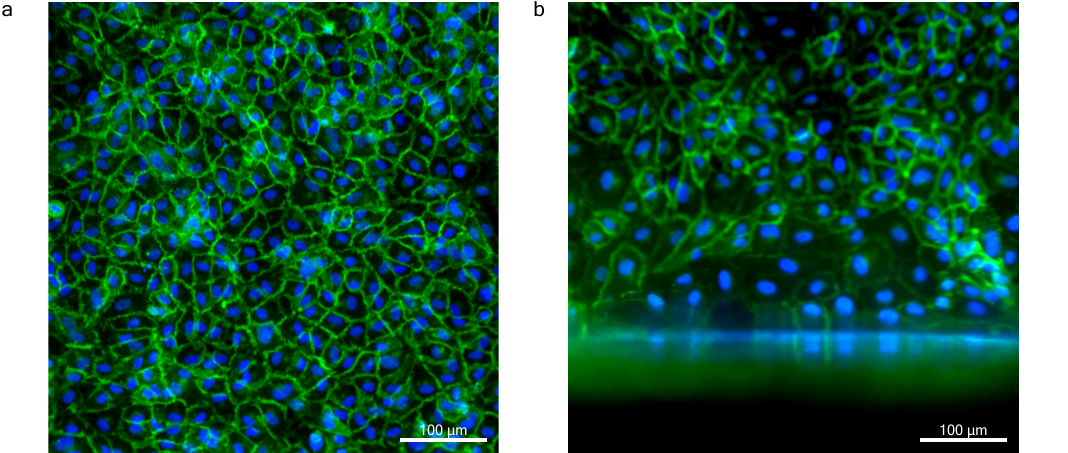}
    \caption{Fluorescence images of primary human umbilical vein endothelial cells grown on a plastic substrate.
    Cells are stained with DAPI (blue) and CD31  antibody (green).
    Confluence of the cell monolayer (a) in the center and (b) at the edge of the membrane.}
    \label{fig:Cellmonolayer}
\end{figure}

\begin{figure}[!ht] 
    \centering
        \includegraphics[width=0.5\columnwidth]{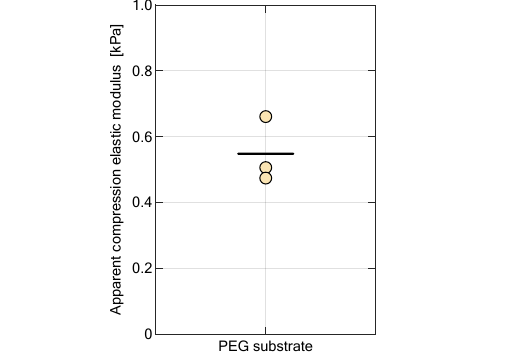}
    \caption{Apparent compression elastic modulus of PEG hydrogel samples measured by atomic force microscopy. 
    Each data point represents the average of median values of the elastic modulus measured at 4–5 different locations of a hydrogel sample, with each location sampled on a $5 \times5$ points grid.}
    \label{fig:PEGmodulus}
\end{figure}

\clearpage

\section*{Supplementary information}

\subsection*{Supplementary Videos description}

\noindent\textbf{Supplementary Video 1}

\noindent Response of a \SI{3}{\micro\meter}-radius microbubble in contact with an endothelial cell to an ultrasound pulse ($f = \SI{1}{\mega\hertz}$, $p_{\rm a} = \SI{60}{\kilo\pascal}$, 20 cycles), captured from a side-view perspective. 
The amplitude of the applied ultrasound pressure is insufficient to induce the formation of cyclic piercing microjets from the microbubble.
The field of view is \SI[product-units = single]{40 x 40}{\micro\metre}.
This video corresponds to the image sequence shown in Fig.~\ref{fig:Fig1}c.

\vspace{2mm}

\noindent\textbf{Supplementary Video 2}

\noindent Response of the same \SI{3}{\micro\meter}-radius microbubble in contact with an endothelial cell to a more intense ultrasound pulse ($f = \SI{1}{\mega\hertz}$, $p_{\rm a} = \SI{160}{\kilo\pascal}$, 20 cycles), captured from a side-view perspective. 
The higher amplitude of the applied ultrasound pressure leads to the formation of cyclic piercing microjets directed towards the cell, resulting in cell membrane poration and drug uptake.
The bubble motion also results in the formation of a transendothelial tunnel.
The field of view is \SI[product-units = single]{40 x 40}{\micro\metre}.
This video corresponds to the image sequence shown in Fig.~\ref{fig:Fig1}d.

\vspace{2mm}

\noindent\textbf{Supplementary Video 3}

\noindent Response of a \SI{3}{\micro\meter}-radius microbubble in contact with an endothelial cell to an ultrasound pulse ($f = \SI{1}{\mega\hertz}$, $p_{\rm a} = \SI{175}{\kilo\pascal}$, 20 cycles), captured from a side-view perspective. 
In this instance, the bubble generates cyclic piercing microjets that facilitate drug uptake, but the motion of the bubble does not result in the formation of a transendothelial tunnel.
The field of view is \SI[product-units = single]{40 x 40}{\micro\metre}.
This video corresponds to the image sequence shown in Extended Data Fig.~\ref{fig:FigNoTunnel}.

\vspace{2mm}

\noindent\textbf{Supplementary Video 4}

\noindent Response of a \SI{2.3}{\micro\meter}-radius microbubble in contact with a PEG substrate to a very intense ultrasound pulse ($f = \SI{1}{\mega\hertz}$, $p_{\rm a} = \SI{2.7}{\mega\pascal}$, 20 cycles), captured from a side-view perspective. 
The bubble generates a single transient inertial jet followed by bubble fragmentation.
The field of view is \SI[product-units = single]{40 x 40}{\micro\metre}.
This video corresponds to the image sequence shown in Extended Data Fig.~\ref{fig:InertialJetting}.

\vspace{2mm}

\noindent\textbf{Supplementary Video 5}

\noindent Response of a \SI{2.9}{\micro\meter}-radius microbubble in contact with an endothelial cell to an ultrasound pulse ($f = \SI{1}{\mega\hertz}$, $p_{\rm a} = \SI{150}{\kilo\pascal}$, 20 cycles), captured from a side-view perspective. 
Cyclic jetting is driven by a shape mode with wavenumber $l = 1$, as the bubble displays an alternate body motion.
The field of view is \SI[product-units = single]{40 x 40}{\micro\metre}.
This video corresponds to the image sequence shown in Extended Data Fig.~\ref{fig:Shapemode_Cells}a.

\vspace{2mm}

\noindent\textbf{Supplementary Video 6}

\noindent Response of a \SI{3.8}{\micro\meter}-radius microbubble in contact with an endothelial cell to an ultrasound pulse ($f = \SI{1}{\mega\hertz}$, $p_{\rm a} = \SI{80}{\kilo\pascal}$, 20 cycles), captured from a side-view perspective. 
Cyclic jetting is driven by a shape mode with wavenumber $l = 2$, as the bubble alternately takes on a prolate and oblate shape.
The field of view is \SI[product-units = single]{40 x 40}{\micro\metre}.
This video corresponds to the image sequence shown in Extended Data Fig.~\ref{fig:Shapemode_Cells}b.

\newpage
\bibliography{main}

@article{Lajoinie2016,
    title = {{In vitro methods to study bubble-cell interactions: Fundamentals and therapeutic applications}},
    year = {2016},
    journal = {Biomicrofluidics},
    author = {Lajoinie, G. and De Cock, I. and Coussios, C. C. and Lentacker, I. and Le Gac, S. and Stride, E. and Versluis, M.},
    number = {1},
    pages = {1--25},
    volume = {10},
    doi = {10.1063/1.4940429},
    issn = {19321058}
}

@article{Vos2011,
    title = {{Nonspherical Shape Oscillations of Coated Microbubbles in Contact With a Wall}},
    year = {2011},
    journal = {Ultrasound in Medicine and Biology},
    author = {Vos, H. J. and Dollet, B. and Versluis, M. and De Jong, N.},
    number = {6},
    pages = {935--948},
    volume = {37},
    doi = {10.1016/j.ultrasmedbio.2011.02.013},
    issn = {03015629},
    pmid = {21601137}
}

@article{Dimcevski2016ACancer,
    title = {{A human clinical trial using ultrasound and microbubbles to enhance gemcitabine treatment of inoperable pancreatic cancer}},
    year = {2016},
    journal = {Journal of Controlled Release},
    author = {Dimcevski, G. and Kotopoulis, S. and Bj{\aa}nes, T. and Hoem, D. and Schj{\o}t, J. and Gjertsen, B. T. and Biermann, M. and Molven, A. and Sorbye, H. and McCormack, E. and Postema, M. and Gilja, O. H.},
    pages = {172--181},
    volume = {243},
    doi = {10.1016/j.jconrel.2016.10.007}
}

@article{Hosseinkhah2012AMicrovessels,
    title = {{A three-dimensional model of an ultrasound contrast agent gas bubble and its mechanical effects on microvessels}},
    year = {2012},
    journal = {Physics in Medicine and Biology},
    author = {Hosseinkhah, N. and Hynynen, K.},
    number = {3},
    volume = {57},
    doi = {10.1088/0031-9155/57/3/785},
    issn = {00319155}
}

@article{Jiang2024AbyssCollisions,
    title = {{Abyss Aerosols: Drop Production from Underwater Bubble Collisions}},
    year = {2024},
    journal = {Physical Review Letters},
    author = {Jiang, X. and Rotily, L. and Villermaux, E. and Wang, X.},
    number = {2},
    volume = {133},
    doi = {10.1103/PhysRevLett.133.024001}
}

@article{Kooiman2014AcousticDelivery,
    title = {{Acoustic behavior of microbubbles and implications for drug delivery}},
    year = {2014},
    journal = {Advanced Drug Delivery Reviews},
    author = {Kooiman, K. and Vos, H. J. and Versluis, M. and De Jong, N.},
    volume = {72},
    doi = {10.1016/j.addr.2014.03.003},
    issn = {18728294}
}

@article{Zhao2005AsymmetricAgents,
    title = {{Asymmetric oscillation of adherent targeted ultrasound contrast agents}},
    year = {2005},
    journal = {Applied Physics Letters},
    author = {Zhao, S. and Ferrara, K. W. and Dayton, P. A.},
    number = {13},
    volume = {87},
    doi = {10.1063/1.2061872},
    issn = {00036951}
}

@article{Helfield2016BiophysicalSonoporation,
    title = {{Biophysical insight into mechanisms of sonoporation}},
    year = {2016},
    journal = {Proceedings of the National Academy of Sciences of the United States of America},
    author = {Helfield, B. and Chen, X. and Watkins, S. C. and Villanueva, F. S.},
    number = {36},
    pages = {9983--9988},
    volume = {113},
    doi = {10.1073/pnas.1606915113}
}

@article{Mettin1997BjerknesField,
    title = {{Bjerknes forces between small cavitation bubbles in a strong acoustic field}},
    year = {1997},
    journal = {Physical Review E},
    author = {Mettin, R. and Akhatov, I. and Parlitz, U. and Ohl, C. D. and Lauterborn, W.},
    number = {3},
    pages = {2924--2931},
    volume = {56},
    doi = {10.1103/PhysRevE.56.2924}
}

@article{Chen2011BloodCavitation,
    title = {{Blood vessel deformations on microsecond time scales by ultrasonic cavitation}},
    year = {2011},
    journal = {Physical Review Letters},
    author = {Chen, H. and Kreider, W. and Brayman, A. A. and Bailey, M. R. and Matula, T. J.},
    number = {3},
    volume = {106},
    doi = {10.1103/PhysRevLett.106.034301},
    issn = {00319007}
}

@article{Mainprize2019Blood-BrainStudy,
    title = {{Blood-Brain Barrier Opening in Primary Brain Tumors with Non-invasive MR-Guided Focused Ultrasound: A Clinical Safety and Feasibility Study}},
    year = {2019},
    journal = {Scientific Reports},
    author = {Mainprize, T. and Lipsman, N. and Huang, Y. and Meng, Y. and Bethune, A. J. and Ironside, S. and Heyn, C. and Alkins, R. and Trudeau, M. and Sahgal, A. and Perry, J. and Hynynen, K.},
    number = {1},
    volume = {9},
    doi = {10.1038/s41598-018-36340-0}
}

@article{Gasca-Salas2021Blood-brainDementia,
    title = {{Blood-brain barrier opening with focused ultrasound in Parkinson’s disease dementia}},
    year = {2021},
    journal = {Nature Communications},
    author = {Gasca-Salas, C. and Fern{\'{a}}ndez-Rodr{\'{i}}guez, B. and Pineda-Pardo, J. A. and Rodr{\'{i}}guez-Rojas, R. and Obeso, I. and Hern{\'{a}}ndez-Fern{\'{a}}ndez, F. and del {\'{A}}lamo, M. and Mata, D. and Guida, P. and Ord{\'{a}}s-Bandera, C. and Rachmilevitch, I. and Obeso, J. A.},
    number = {1},
    volume = {12},
    doi = {10.1038/s41467-021-21022-9}
}

@article{Lipsman2018BloodbrainUltrasound,
    title = {{Blood–brain barrier opening in Alzheimer’s disease using MR-guided focused ultrasound}},
    year = {2018},
    journal = {Nature Communications},
    author = {Lipsman, N. and Meng, Y. and Bethune, A. J. and Huang, Y. and Lam, B. and Masellis, M. and Herrmann, N. and Heyn, C. and Aubert, I. and Boutet, A. and Hynynen, K. and Black, S. E.},
    number = {1},
    volume = {9},
    doi = {10.1038/s41467-018-04529-6}
}

@article{Prosperetti1982BubbleResults,
    title = {{Bubble dynamics: a review and some recent results}},
    year = {1982},
    journal = {Applied Scientific Research},
    author = {Prosperetti, A.},
    number = {1},
    pages = {145--164},
    volume = {38},
    doi = {10.1007/BF00385945}
}

@article{Longuet-Higgins1983BubblesSurface,
    title = {{Bubbles, breaking waves and hyperbolic jets at a free surface}},
    year = {1983},
    journal = {Journal of Fluid Mechanics},
    author = {Longuet-Higgins, M. S.},
    volume = {127},
    doi = {10.1017/S0022112083002645},
    issn = {14697645}
}

@article{Field2012CavitationThresholds,
    title = {{Cavitation in impacted drops and jets and the effect on erosion damage thresholds}},
    year = {2012},
    journal = {Wear},
    author = {Field, J. E. and Camus, J. J. and Tinguely, M. and Obreschkow, D. and Farhat, M.},
    volume = {290-291},
    doi = {10.1016/j.wear.2012.03.006},
    issn = {00431648}
}

@article{Sheikov2004CellularMicrobubbles,
    title = {{Cellular mechanisms of the blood-brain barrier opening induced by ultrasound in presence of microbubbles}},
    year = {2004},
    journal = {Ultrasound in Medicine and Biology},
    author = {Sheikov, N. and McDannold, N. and Vykhodtseva, N. and Jolesz, F. and Hynynen, K.},
    number = {7},
    volume = {30},
    doi = {10.1016/j.ultrasmedbio.2004.04.010},
    issn = {03015629}
}

@article{Carpentier2016ClinicalUltrasound,
    title = {{Clinical trial of blood-brain barrier disruption by pulsed ultrasound}},
    year = {2016},
    journal = {Science Translational Medicine},
    author = {Carpentier, A. and Canney, M. and Vignot, A. and Reina, V. and Beccaria, K. and Horodyckid, C. and Karachi, C. and Leclercq, D. and Lafon, C. and Chapelon, J.-Y. and Delattre, J.-Y. and Idbaih, A.},
    number = {343},
    volume = {8},
    doi = {10.1126/scitranslmed.aaf6086}
}

@article{Marmottant2003ControlledBubbles,
    title = {{Controlled vesicle deformation and lysis by single oscillating bubbles}},
    year = {2003},
    journal = {Nature},
    author = {Marmottant, P. and Hilgenfeldt, S.},
    number = {6936},
    pages = {153--156},
    volume = {423},
    doi = {10.1038/nature01613}
}

@article{vanElburg2023DependenceStudy,
    title = {{Dependence of sonoporation efficiency on microbubble size: An in vitro monodisperse microbubble study}},
    year = {2023},
    journal = {Journal of Controlled Release},
    author = {van Elburg, B. and Deprez, J. and van den Broek, M. and De Smedt, S. C. and Versluis, M. and Lajoinie, G. and Lentacker, I. and Segers, T.},
    volume = {363},
    doi = {10.1016/j.jconrel.2023.09.047},
    issn = {18734995}
}

@article{Guedra2017DynamicsThreshold,
    title = {{Dynamics of nonspherical microbubble oscillations above instability threshold}},
    year = {2017},
    journal = {Physical Review E},
    author = {Gu{\'{e}}dra, M. and Cleve, S. and Mauger, C. and Blanc-Benon, P. and Inserra, C.},
    number = {6},
    volume = {96},
    doi = {10.1103/PhysRevE.96.063104}
}

@article{Mitchell2021EngineeringDelivery,
    title = {{Engineering precision nanoparticles for drug delivery}},
    year = {2021},
    journal = {Nature Reviews Drug Discovery},
    author = {Mitchell, M. J. and Billingsley, M. M. and Haley, R. M. and Wechsler, M. E. and Peppas, N. A. and Langer, R.},
    number = {2},
    pages = {101--124},
    volume = {20},
    doi = {10.1038/s41573-020-0090-8}
}

@article{Cook1928ErosionWater-hammer,
    title = {{Erosion by water-hammer}},
    year = {1928},
    journal = {Proceedings of the Royal Society of London. Series A},
    author = {Cook, S. S.},
    number = {783},
    volume = {119},
    doi = {10.1098/rspa.1928.0107},
    issn = {0950-1207}
}

@article{Guedra2016ExperimentalMicrobubbles,
    title = {{Experimental evidence of nonlinear mode coupling between spherical and nonspherical oscillations of microbubbles}},
    year = {2016},
    journal = {Physical Review E},
    author = {Gu{\'{e}}dra, M. and Inserra, C. and Mauger, C. and Gilles, B.},
    number = {5},
    volume = {94},
    doi = {10.1103/PhysRevE.94.053115}
}

@article{Ebo-Adou2019FaradaySimulation,
    title = {{Faraday instability on a sphere: Numerical simulation}},
    year = {2019},
    journal = {Journal of Fluid Mechanics},
    author = {Ebo-Adou, A. and Tuckerman, L. S. and Shin, S. and Chergui, J. and Juric, D.},
    pages = {433--459},
    volume = {870},
    doi = {10.1017/jfm.2019.252}
}

@article{Lechner2019FastStudy,
    title = {{Fast, thin jets from bubbles expanding and collapsing in extreme vicinity to a solid boundary: A numerical study}},
    year = {2019},
    journal = {Physical Review Fluids},
    author = {Lechner, C. and Lauterborn, W. and Koch, M. and Mettin, R.},
    number = {2},
    volume = {4},
    doi = {10.1103/PhysRevFluids.4.021601},
    issn = {2469990X}
}

@article{Abrahao2019First-in-humanUltrasound,
    title = {{First-in-human trial of blood–brain barrier opening in amyotrophic lateral sclerosis using MR-guided focused ultrasound}},
    year = {2019},
    journal = {Nature Communications},
    author = {Abrahao, A. and Meng, Y. and Llinas, M. and Huang, Y. and Hamani, C. and Mainprize, T. and Aubert, I. and Heyn, C. and Black, S.E. and Hynynen, K. and Lipsman, N. and Zinman, L.},
    number = {1},
    volume = {10},
    doi = {10.1038/s41467-019-12426-9}
}

@article{Gekle2010GenerationFormation,
    title = {{Generation and breakup of Worthington jets after cavity collapse. Part 1. Jet formation}},
    year = {2010},
    journal = {Journal of Fluid Mechanics},
    author = {Gekle, S. and Gordillo, J. M.},
    volume = {663},
    doi = {10.1017/S0022112010003526},
    issn = {14697645}
}

@article{Gekle2009High-speedImpact,
    title = {{High-speed jet formation after solid object impact}},
    year = {2009},
    journal = {Physical Review Letters},
    author = {Gekle, S. and Gordillo, J. M. and Van Der Meer, D. and Lohse, D.},
    number = {3},
    volume = {102},
    doi = {10.1103/PhysRevLett.102.034502},
    issn = {00319007}
}

@book{Lamb1932Hydrodynamics,
    title = {{Hydrodynamics}},
    year = {1932},
    booktitle = {Hydrodynamics by Horace Lamb ...},
    author = {Lamb, H.},
    publisher = {Cambridge University Press},
    doi = {10.5962/bhl.title.18729}
}

@article{Anbarafshan2024InModel,
    title = {{In vivo high-speed microscopy of microbubbles in the chorioallantoic membrane model}},
    year = {2024},
    journal = {Theranostics},
    author = {Anbarafshan, R. and Pellow, C. and Kiezun, K. and Leong, H. and Goertz, D. E.},
    number = {5},
    volume = {14},
    doi = {10.7150/thno.91232},
    issn = {18387640}
}

@article{Bulycheva2024InteractionBoundary,
    title = {{Interaction of ultrasonically driven bubble with a soft tissue-like boundary}},
    year = {2024},
    journal = {Ultrasonics},
    author = {Bulycheva, V. and Kolios, M.C. and Karshafian, R.},
    volume = {142},
    doi = {10.1016/j.ultras.2024.107374}
}

@article{Beekers2022InternalizationTunnels,
    title = {{Internalization of targeted microbubbles by endothelial cells and drug delivery by pores and tunnels}},
    year = {2022},
    journal = {Journal of Controlled Release},
    author = {Beekers, I. and Langeveld, S. A. G. and Meijlink, B. and van der Steen, A. F. W. and de Jong, N. and Verweij, M. D. and Kooiman, K.},
    pages = {460--475},
    volume = {347},
    doi = {10.1016/j.jconrel.2022.05.008}
}

@article{Duchemin2002JetSurface,
    title = {{Jet formation in bubbles bursting at a free surface}},
    year = {2002},
    journal = {Physics of Fluids},
    author = {Duchemin, L. and Popinet, S. and Josserand, C. and Zaleski, S.},
    number = {9},
    volume = {14},
    doi = {10.1063/1.1494072},
    issn = {10706631}
}

@article{Ferguson2018KinaseAhead,
    title = {{Kinase inhibitors: The road ahead}},
    year = {2018},
    journal = {Nature Reviews Drug Discovery},
    author = {Ferguson, F. M. and Gray, N. S.},
    number = {5},
    pages = {353--376},
    volume = {17},
    doi = {10.1038/nrd.2018.21}
}

@article{Pereno2018LayeredEffects,
    title = {{Layered acoustofluidic resonators for the simultaneous optical and acoustic characterisation of cavitation dynamics, microstreaming, and biological effects}},
    year = {2018},
    journal = {Biomicrofluidics},
    author = {Pereno, V. and Aron, M. and Vince, O. and Mannaris, C. and Seth, A. and De Saint Victor, M. and Lajoinie, G. and Versluis, M. and Coussios, C. and Carugo, D. and Stride, E.},
    number = {3},
    volume = {12},
    doi = {10.1063/1.5023729},
    issn = {19321058}
}

@article{Yeh2012Matrix9,
    title = {{Matrix Stiffness Regulates Endothelial Cell Proliferation through Septin 9}},
    year = {2012},
    journal = {PLoS ONE},
    author = {Yeh, Y. T. and Hur, S. S. and Chang, J. and Wang, K. C. and Chiu, J. J. and Li, Y. S. and Chien, S.},
    number = {10},
    volume = {7},
    doi = {10.1371/journal.pone.0046889},
    issn = {19326203}
}

@article{Budday2015MechanicalIndentation,
    title = {{Mechanical properties of gray and white matter brain tissue by indentation}},
    year = {2015},
    journal = {Journal of the Mechanical Behavior of Biomedical Materials},
    author = {Budday, S. and Nay, R. and de Rooij, R. and Steinmann, P. and Wyrobek, T. and Ovaert, T. C. and Kuhl, E.},
    volume = {46},
    doi = {10.1016/j.jmbbm.2015.02.024},
    issn = {18780180}
}

@article{Prentice2005MembraneCavitation,
    title = {{Membrane disruption by optically controlled microbubble cavitation}},
    year = {2005},
    journal = {Nature Physics},
    author = {Prentice, P. and Cuschieri, A. and Dholakia, K. and Prausnitz, M. and Campbell, P.},
    number = {2},
    pages = {107--110},
    volume = {1},
    doi = {10.1038/nphys148}
}

@article{Bezer2023MicrobubbleMicrovessels,
    title = {{Microbubble dynamics in brain microvessels}},
    year = {2023},
    journal = {bioRxiv},
    author = {Bezer, J. H. and Prentice, P. and Lim, W. and Chang, K. and Morse, S. V. and Christensen-Jeffries, K. and Rowlands, C. J. and Kozlov, A. S. and Choi, J. J.}
}

@article{Meng2021MR-guidedMetastases,
    title = {{MR-guided focused ultrasound enhances delivery of trastuzumab to Her2-positive brain metastases}},
    year = {2021},
    journal = {Science Translational Medicine},
    author = {Meng, Y. and Reilly, R. M. and Pezo, R. C. and Trudeau, M. and Sahgal, A. and Singnurkar, A. and Perry, J. and Myrehaug, S. and Pople, C. B. and Davidson, B. and Hynynen, K. and Lipsman, N.},
    number = {615},
    volume = {13},
    doi = {10.1126/scitranslmed.abj4011}
}

@article{Chen2021Neuronavigation-guidedTumors,
    title = {{Neuronavigation-guided focused ultrasound for transcranial blood-brain barrier opening and immunostimulation in brain tumors}},
    year = {2021},
    journal = {Science Advances},
    author = {Chen, K.-T. and Chai, W.-Y. and Lin, Y.-J. and Lin, C.-J. and Chen, P.-Y. and Tsai, H.-C. and Huang, C.-Y. and Kuo, J.S. and Liu, H.-L. and Wei, K.-C.},
    number = {6},
    volume = {7},
    doi = {10.1126/sciadv.abd0772}
}

@article{Lajoinie2018Non-sphericalMicrobubbles,
    title = {{Non-spherical oscillations drive the ultrasound-mediated release from targeted microbubbles}},
    year = {2018},
    journal = {Communications Physics},
    author = {Lajoinie, G. and Luan, Y. and Gelderblom, E. and Dollet, B. and Mastik, F. and Dewitte, H. and Lentacker, I. and de Jong, N. and Versluis, M.},
    number = {1},
    volume = {1},
    doi = {10.1038/s42005-018-0020-9},
    issn = {23993650}
}

@article{Rezai2020NoninvasiveUltrasound,
    title = {{Noninvasive hippocampal blood-brain barrier opening in Alzheimer’s disease with focused ultrasound}},
    year = {2020},
    journal = {Proceedings of the National Academy of Sciences of the United States of America},
    author = {Rezai, A. R. and Ranjan, M. and D’Haese, P.-F. and Haut, M. W. and Carpenter, J. and Najib, U. and Mehta, R. I. and Chazen, J. L. and Zibly, Z. and Yates, J. R. and Hodder, S. L. and Kaplitt, M.},
    number = {17},
    pages = {9180--9182},
    volume = {117},
    doi = {10.1073/pnas.2002571117}
}

@article{Riahi1984NonlinearShell,
    title = {{Nonlinear Convection in a Spherical Shell}},
    year = {1984},
    journal = {Journal of the Physical Society of Japan},
    author = {Riahi, N.},
    number = {8},
    volume = {53},
    doi = {10.1143/JPSJ.53.2506},
    issn = {13474073}
}

@article{Dollet2008NonsphericalMicrobubbles,
    title = {{Nonspherical Oscillations of Ultrasound Contrast Agent Microbubbles}},
    year = {2008},
    journal = {Ultrasound in Medicine and Biology},
    author = {Dollet, B. and van der Meer, S. M. and Garbin, V. and de Jong, N. and Lohse, D. and Versluis, M.},
    number = {9},
    volume = {34},
    doi = {10.1016/j.ultrasmedbio.2008.01.020},
    issn = {03015629}
}

@article{Stride2019NucleationDelivery,
    title = {{Nucleation, mapping and control of cavitation for drug delivery}},
    year = {2019},
    journal = {Nature Reviews Physics},
    author = {Stride, E. and Coussios, C.},
    number = {8},
    volume = {1},
    doi = {10.1038/s42254-019-0074-y},
    issn = {25225820}
}

@article{Beekers2020OpeningSonoporation,
    title = {{Opening of endothelial cell–cell contacts due to sonoporation}},
    year = {2020},
    journal = {Journal of Controlled Release},
    author = {Beekers, I. and Vegter, M. and Lattwein, K. R. and Mastik, F. and Beurskens, R. and van der Steen, A. F. W. and de Jong, N. and Verweij, M. D. and Kooiman, K.},
    pages = {426--438},
    volume = {322},
    doi = {10.1016/j.jconrel.2020.03.038}
}

@article{Konofagou2012OptimizationOpening,
    title = {{Optimization of the ultrasound-induced blood-brain barrier opening}},
    year = {2012},
    journal = {Theranostics},
    author = {Konofagou, E. E.},
    number = {12},
    pages = {1223--1237},
    volume = {2},
    doi = {10.7150/thno.5576}
}

@article{Busse1975PatternsShells,
    title = {{Patterns of convection in spherical shells}},
    year = {1975},
    journal = {Journal of Fluid Mechanics},
    author = {Busse, F. H.},
    number = {1},
    volume = {72},
    doi = {10.1017/S0022112075002947},
    issn = {14697645}
}

@article{Busse1982Patterns2,
    title = {{Patterns of convection in spherical shells. Part 2}},
    year = {1982},
    journal = {Journal of Fluid Mechanics},
    author = {Busse, F. H. and Riahi, N.},
    volume = {123},
    doi = {10.1017/S0022112082003061},
    issn = {14697645}
}

@article{Kientzler1954PhotographicSurface,
    title = {{Photographic Investigation of the Projection of Droplets by Bubbles Bursting at a Water Surface}},
    year = {1954},
    journal = {Tellus A: Dynamic Meteorology and Oceanography},
    author = {Kientzler, C. F. and Arons, A. B.},
    number = {1},
    volume = {6},
    doi = {10.3402/tellusa.v6i1.8717}
}

@article{Blanco2015PrinciplesDelivery,
    title = {{Principles of nanoparticle design for overcoming biological barriers to drug delivery}},
    year = {2015},
    journal = {Nature Biotechnology},
    author = {Blanco, E. and Shen, H. and Ferrari, M.},
    number = {9},
    volume = {33},
    doi = {10.1038/nbt.3330},
    issn = {15461696}
}

@article{Sonabend2023RepeatedTrial,
    title = {{Repeated blood–brain barrier opening with an implantable ultrasound device for delivery of albumin-bound paclitaxel in patients with recurrent glioblastoma: a phase 1 trial}},
    year = {2023},
    journal = {The Lancet Oncology},
    author = {Sonabend, A. M. and Gould, A. and Amidei, C. and Ward, R. and Schmidt, K. A. and Zhang, D. Y. and Gomez, C. and Bebawy, J. F. and Liu, B. P. and Bouchoux, G. and Canney, M. and Stupp, R.},
    number = {5},
    pages = {509--522},
    volume = {24},
    doi = {10.1016/S1470-2045(23)00112-2}
}

@article{Idbaih2019SafetyGlioblastoma,
    title = {{Safety and feasibility of repeated and transient blood-brain barrier disruption by pulsed ultrasound in patients with recurrent glioblastoma}},
    year = {2019},
    journal = {Clinical Cancer Research},
    author = {Idbaih, A. and Canney, M. and Belin, L. and Desseaux, C. and Vignot, A. and Bouchoux, G. and Asquier, N. and Law-Ye, B. and Leclercq, D. and Bissery, A. and Delattre, J.-Y. and Carpentier, A.},
    number = {13},
    pages = {3793--3801},
    volume = {25},
    doi = {10.1158/1078-0432.CCR-18-3643}
}

@article{Poulichet2017ShapeExpulsion,
    title = {{Shape oscillations of particle-coated bubbles and directional particle expulsion}},
    year = {2017},
    journal = {Soft Matter},
    author = {Poulichet, V. and Huerre, A. and Garbin, V.},
    number = {1},
    volume = {13},
    doi = {10.1039/C6SM01603K},
    issn = {17446848}
}

@article{Rooney1972ShearEffects,
    title = {{Shear as a Mechanism for Sonically Induced Biological Effects}},
    year = {1972},
    journal = {Journal of the Acoustical Society of America},
    author = {Rooney, J. A.},
    number = {6B},
    pages = {1718--1724},
    volume = {52},
    doi = {10.1121/1.1913306}
}

@article{Cattaneo2023ShellMicrobubbles,
    title = {{Shell viscosity estimation of lipid-coated microbubbles}},
    year = {2023},
    journal = {Soft Matter},
    author = {Cattaneo, M. and Supponen, O.},
    number = {31},
    pages = {5925--5941},
    volume = {19},
    doi = {10.1039/d3sm00871a}
}

@article{Thoroddsen2018SingularCraters,
    title = {{Singular jets during the collapse of drop-impact craters}},
    year = {2018},
    journal = {Journal of Fluid Mechanics},
    author = {Thoroddsen, S. T. and Takehara, K. and Nguyen, H. D. and Etoh, T. G.},
    volume = {848},
    doi = {10.1017/jfm.2018.435},
    issn = {14697645}
}

@article{Zeff2000SingularitySurface,
    title = {{Singularity dynamics in curvature collapse and jet eruption on a fluid surface}},
    year = {2000},
    journal = {Nature},
    author = {Zeff, B. W. and Kleber, B. and Fineberg, J. and Lathrop, D. P.},
    number = {6768},
    pages = {401--404},
    volume = {403},
    doi = {10.1038/35000151}
}

@article{Mathias2019SonothrombolysisIntervention,
    title = {{Sonothrombolysis in ST-Segment Elevation Myocardial Infarction Treated With Primary Percutaneous Coronary Intervention}},
    year = {2019},
    journal = {Journal of the American College of Cardiology},
    author = {Mathias, W. and Tsutsui, J. M. and Tavares, B. G. and Fava, A. M. and Aguiar, M. O. D. and Borges, B. C. and Oliveira, M. T. and Soeiro, A. and Nicolau, J. C. and Ribeiro, H. B. and Kalil Filho, R. and Porter, T. R.},
    number = {22},
    pages = {2832--2842},
    volume = {73},
    doi = {10.1016/j.jacc.2019.03.006}
}

@article{Fan2012SpatiotemporallySonoporation,
    title = {{Spatiotemporally controlled single cell sonoporation}},
    year = {2012},
    journal = {Proceedings of the National Academy of Sciences of the United States of America},
    author = {Fan, Z. and Liu, H. and Mayer, M. and Deng, C. X.},
    number = {41},
    pages = {16486--16491},
    volume = {109},
    doi = {10.1073/pnas.1208198109}
}

@article{Chossat1991Steady-State03-Symmetry,
    title = {{Steady-State bifurcation with 0(3)-Symmetry}},
    year = {1991},
    journal = {Archive for Rational Mechanics and Analysis},
    author = {Chossat, P. and Lauterbach, R. and Melbourne, I.},
    number = {4},
    volume = {113},
    doi = {10.1007/BF00374697},
    issn = {00039527}
}

@article{Reuter2021SupersonicBubbles,
    title = {{Supersonic needle-jet generation with single cavitation bubbles}},
    year = {2021},
    journal = {Applied Physics Letters},
    author = {Reuter, F. and Ohl, C. D.},
    number = {13},
    volume = {118},
    doi = {10.1063/5.0045705},
    issn = {00036951}
}

@article{Prabowo2011SurfaceBubbles,
    title = {{Surface oscillation and jetting from surface attached acoustic driven bubbles}},
    year = {2011},
    journal = {Ultrasonics Sonochemistry},
    author = {Prabowo, F. and Ohl, C. D.},
    number = {1},
    volume = {18},
    doi = {10.1016/j.ultsonch.2010.07.013},
    issn = {13504177}
}

@article{Crum1979SurfaceBubbles,
    title = {{Surface oscillations and jet development in pulsating bubbles}},
    year = {1979},
    journal = {J. Phys. Colloques},
    author = {Crum, L. A.},
    number = {C8},
    month = {11},
    pages = {8--285},
    volume = {40},
}

@article{Manzari2021TargetedMedicines,
    title = {{Targeted drug delivery strategies for precision medicines}},
    year = {2021},
    journal = {Nature Reviews Materials},
    author = {Manzari, M. T. and Shamay, Y. and Kiguchi, H. and Rosen, N. and Scaltriti, M. and Heller, D. A.},
    number = {4},
    volume = {6},
    doi = {10.1038/s41578-020-00269-6},
    issn = {20588437}
}

@article{Meng2021TechnicalUltrasound,
    title = {{Technical Principles and Clinical Workflow of Transcranial MR-Guided Focused Ultrasound}},
    year = {2021},
    journal = {Stereotactic and Functional Neurosurgery},
    author = {Meng, Y. and Jones, R. M. and Davidson, B. and Huang, Y. and Pople, C. B. and Surendrakumar, S. and Hamani, C. and Hynynen, K. and Lipsman, N.},
    number = {4},
    pages = {329--342},
    volume = {99},
    doi = {10.1159/000512111}
}

@article{Zhou2009TheMembrane,
    title = {{The Size of Sonoporation Pores on the Cell Membrane}},
    year = {2009},
    journal = {Ultrasound in Medicine and Biology},
    author = {Zhou, Y. and Kumon, R. E. and Cui, J. and Deng, C. X.},
    number = {10},
    volume = {35},
    doi = {10.1016/j.ultrasmedbio.2009.05.012},
    issn = {03015629}
}

@article{Gordillo2023TheoryJets,
    title = {{Theory of the jets ejected after the inertial collapse of cavities with applications to bubble bursting jets}},
    year = {2023},
    journal = {Physical Review Fluids},
    author = {Gordillo, J. M. and Blanco-Rodr{\'{i}}guez, F. J.},
    number = {7},
    volume = {8},
    doi = {10.1103/PhysRevFluids.8.073606},
    issn = {2469990X}
}

@article{Helfield2020TransendothelialSonoporation,
    title = {{Transendothelial Perforations and the Sphere of Influence of Single-Site Sonoporation}},
    year = {2020},
    journal = {Ultrasound in Medicine {\&} Biology},
    author = {Helfield, B. and Chen, X. and Watkins, S. C. and Villanueva, F. S.},
    number = {7},
    pages = {1686--1697},
    volume = {46},
    doi = {https://doi.org/10.1016/j.ultrasmedbio.2020.02.017},
    issn = {0301-5629}
}

@article{Bao1997TransfectionVitro,
    title = {{Transfection of a reporter plasmid into cultured cells by sonoporation in vitro}},
    year = {1997},
    journal = {Ultrasound in Medicine and Biology},
    author = {Bao, S. and Thrall, B. D. and Miller, D. L.},
    number = {6},
    volume = {23},
    doi = {10.1016/S0301-5629(97)00025-2},
    issn = {03015629}
}

@article{Curtiss2013UltrasonicLayer,
    title = {{Ultrasonic cavitation near a tissue layer}},
    year = {2013},
    journal = {Journal of Fluid Mechanics},
    author = {Curtiss, G. A. and Leppinen, D. M. and Wang, Q. X. and Blake, J. R.},
    volume = {730},
    doi = {10.1017/jfm.2013.341},
    issn = {14697645}
}

@article{DeCock2015UltrasoundEndocytosis,
    title = {{Ultrasound and microbubble mediated drug delivery: acoustic pressure as determinant for uptake via membrane pores or endocytosis}},
    year = {2015},
    journal = {Journal of Controlled Release},
    author = {De Cock, I. and Zagato, E. and Braeckmans, K. and Luan, Y. and de Jong, N. and De Smedt, S. C. and Lentacker, I.},
    volume = {197},
    doi = {10.1016/j.jconrel.2014.10.031},
    issn = {18734995}
}

@article{Kopechek2019UltrasoundFunction,
    title = {{Ultrasound and microbubble-targeted delivery of a microRNA inhibitor to the heart suppresses cardiac hypertrophy and preserves cardiac function}},
    year = {2019},
    journal = {Theranostics},
    author = {Kopechek, J. A. and McTiernan, C. F. and Chen, X. and Zhu, J. and Mburu, M. and Feroze, R. and Whitehurst, D. A. and Lavery, L. and Cyriac, J. and Villanueva, F. S.},
    number = {23},
    pages = {7088--7098},
    volume = {9},
    doi = {10.7150/thno.34895}
}

@article{Versluis2020UltrasoundReview,
    title = {{Ultrasound Contrast Agent Modeling: A Review}},
    year = {2020},
    journal = {Ultrasound in Medicine and Biology},
    author = {Versluis, M. and Stride, E. and Lajoinie, G. and Dollet, B. and Segers, T.},
    number = {9},
    pages = {2117--2144},
    volume = {46},
    doi = {10.1016/j.ultrasmedbio.2020.04.014}
}

@article{Kooiman2020Ultrasound-ResponsiveDelivery,
    title = {{Ultrasound-Responsive Cavitation Nuclei for Therapy and Drug Delivery}},
    year = {2020},
    journal = {Ultrasound in Medicine and Biology},
    author = {Kooiman, K. and Roovers, S. and Langeveld, S. A. G. and Kleven, R. T. and Dewitte, H. and O'Reilly, M. A. and Escoffre, J. M. and Bouakaz, A. and Verweij, M. D. and Hynynen, K. and Lentacker, I. and Stride, E. and Holland, C. K.},
    number = {6},
    volume = {46},
    doi = {10.1016/j.ultrasmedbio.2020.01.002},
    issn = {1879291X}
}

@article{Shakya2024Ultrasound-responsiveDelivery,
    title = {{Ultrasound-responsive microbubbles and nanodroplets: A pathway to targeted drug delivery}},
    year = {2024},
    journal = {Advanced Drug Delivery Reviews},
    author = {Shakya, G. and Cattaneo, M. and Guerriero, G. and Prasanna, A. and Fiorini, S. and Supponen, O.},
    volume = {206},
    doi = {10.1016/j.addr.2023.115178}
}

@article{Obreschkow2011UniversalBubbles,
    title = {{Universal scaling law for jets of collapsing bubbles}},
    year = {2011},
    journal = {Physical Review Letters},
    author = {Obreschkow, D. and Tinguely, M. and Dorsaz, N. and Kobel, P. and De Bosset, A. and Farhat, M.},
    number = {20},
    volume = {107},
    doi = {10.1103/PhysRevLett.107.204501}
}

@article{deHaller1933UntersuchungenKorrosionen,
    title = {{Untersuchungen {\"{u}}ber die durch kavitation hervorgerufenen korrosionen}},
    year = {1933},
    journal = {Schweizerische Bauzeitung},
    author = {de Haller, P.},
    volume = {101}
}

@article{Eisenbrey2021US-triggeredTrial,
    title = {{US-triggered Microbubble Destruction for Augmenting Hepatocellular Carcinoma Response to Transarterial Radioembolization: A Randomized Pilot Clinical Trial}},
    year = {2021},
    journal = {Radiology},
    author = {Eisenbrey, J. R. and Forsberg, F. and Wessner, C. E. and Delaney, L. J. and Bradigan, K. and Gummadi, S. and Tantawi, M. and Lyshchik, A. and O'Kane, P. and Liu, J.-B. and Shamimi-Noori, S. and Shaw, C. M.},
    number = {2},
    pages = {450--457},
    volume = {298},
    doi = {10.1148/RADIOL.2020202321}
}

@article{Worthington1897V.Photography,
    title = {{V. Impact with a liquid surface, studied by the aid of instantaneous photography}},
    year = {1897},
    journal = {Proceedings of the Royal Society of London. Series A},
    author = {Worthington, A. M. and Cole, R. S.},
    volume = {189},
    doi = {10.1098/rsta.1897.0005},
    issn = {0264-3952}
}

@article{Rayleigh1879VI.Jets,
    title = {{VI. On the capillary phenomena of jets}},
    year = {1879},
    journal = {Proceedings of the Royal Society of London},
    author = {Rayleigh, L.},
    number = {196-199},
    volume = {29},
    doi = {10.1098/rspl.1879.0015},
    issn = {0370-1662}
}

@article{vanWamel2006VibratingSonoporation,
    title = {{Vibrating microbubbles poking individual cells: Drug transfer into cells via sonoporation}},
    year = {2006},
    journal = {Journal of Controlled Release},
    author = {van Wamel, A. and Kooiman, K. and Harteveld, M. and Emmer, M. and ten Cate, F. J. and Versluis, M. and de Jong, N.},
    number = {2},
    pages = {149--155},
    volume = {112},
    doi = {10.1016/j.jconrel.2006.02.007}
}

@article{Longuet-Higgins1998ViscousBubble,
    title = {{Viscous streaming from an oscillating spherical bubble}},
    year = {1998},
    journal = {Proceedings of the Royal Society of London. Series A},
    author = {Longuet-Higgins, M. S.},
    number = {1970},
    volume = {454},
    doi = {10.1098/rspa.1998.0183},
    issn = {13645021}
}

@article{Faraday1831XVII.Surfaces,
    title = {{XVII. On a peculiar class of acoustical figures; and on certain forms assumed by groups of particles upon vibrating elastic surfaces}},
    year = {1831},
    journal = {Philosophical Transactions of the Royal Society of London},
    author = {Faraday, M.},
    pages = {299--340},
    volume = {121},
    doi = {10.1098/rstl.1831.0018}
}

@article{Feshitan2009,
    title = {{Microbubble size isolation by differential centrifugation}},
    year = {2009},
    journal = {Journal of Colloid and Interface Science},
    author = {Feshitan, J. A. and Chen, C. C. and Kwan, J. J. and Borden, M. A.},
    number = {2},
    pages = {316--324},
    volume = {329},
    publisher = {Elsevier Inc.},
    doi = {10.1016/j.jcis.2008.09.066},
    issn = {00219797},
    pmid = {18950786}
}

@article{Dudaryeva20213DDeath,
    title = {{3D Confinement Regulates Cell Life and Death}},
    year = {2021},
    journal = {Advanced Functional Materials},
    author = {Dudaryeva, O. Y. and Bucciarelli, A. and Bovone, G. and Huwyler, F. and Jaydev, S. and Broguiere, N. and al-Bayati, M. and L{\"{u}}tolf, M. and Tibbitt, M. W.},
    number = {52},
    volume = {31},
    doi = {10.1002/adfm.202104098},
    issn = {16163028}
}

@article{Marmottant2005ARupture,
    title = {{A model for large amplitude oscillations of coated bubbles accounting for buckling and rupture}},
    year = {2005},
    journal = {Journal of the Acoustical Society of America},
    author = {Marmottant, P. and Van Der Meer, S. and Emmer, M. and Versluis, M. and De Jong, N. and Hilgenfeldt, S. and Lohse, D.},
    number = {6},
    pages = {3499--3505},
    volume = {118},
    doi = {10.1121/1.2109427}
}

@article{Emiroglu2022BuildingDeformations,
    title = {{Building block properties govern granular hydrogel mechanics through contact deformations}},
    year = {2022},
    journal = {Science Advances},
    author = {Emiroglu, D. B. and Bekcic, A. and Dranseikiene, D. and Zhang, X. and Zambelli, T. and deMello, A. J. and Tibbitt, M. W.},
    number = {50},
    volume = {8},
    doi = {10.1126/sciadv.add8570},
    issn = {23752548}
}

@article{Bonfanti2020FractionalMaterials,
    title = {{Fractional viscoelastic models for power-law materials}},
    year = {2020},
    journal = {Soft Matter},
    author = {Bonfanti, A. and Kaplan, J.L. and Charras, G. and Kabla, A.},
    number = {26},
    pages = {6002--6020},
    volume = {16},
    doi = {10.1039/d0sm00354a}
}

@book{Edwards1991InterfacialRheology,
    title = {{Interfacial Transport Processes and Rheology}},
    year = {1991},
    author = {Edwards, D. A. and Brenner, H. and Wasan, D. T.},
    publisher = {Elsevier},
    isbn = {9780750691857},
    doi = {10.1016/C2009-0-26916-9}
}

@article{Hurst2021IntracellularDivision,
    title = {{Intracellular softening and increased viscoelastic fluidity during division}},
    year = {2021},
    journal = {Nature Physics},
    author = {Hurst, S. and Vos, B. E. and Brandt, M. and Betz, T.},
    number = {11},
    pages = {1270--1276},
    volume = {17},
    doi = {10.1038/s41567-021-01368-z}
}

@article{Caputo1967LinearIndependentII,
    title = {{Linear Models of Dissipation whose Q is almost Frequency Independent‐II}},
    year = {1967},
    journal = {Geophysical Journal of the Royal Astronomical Society},
    author = {Caputo, M.},
    number = {5},
    volume = {13},
    doi = {10.1111/j.1365-246X.1967.tb02303.x},
    issn = {1365246X}
}

@article{Kwan2012LipidExchange,
    title = {{Lipid monolayer dilatational mechanics during microbubble gas exchange}},
    year = {2012},
    journal = {Soft Matter},
    author = {Kwan, J. J. and Borden, M. A.},
    number = {17},
    pages = {4756--4766},
    volume = {8},
    doi = {10.1039/c2sm07437k}
}

@article{Zhou2021ModelingBubble,
    title = {{Modeling the thermal behavior of an acoustically driven gas bubble}},
    year = {2021},
    journal = {Journal of the Acoustical Society of America},
    author = {Zhou, G.},
    number = {2},
    pages = {923--933},
    volume = {149},
    doi = {10.1121/10.0003439}
}

@article{Prosperetti1988NonlinearDynamics,
    title = {{Nonlinear bubble dynamics}},
    year = {1988},
    journal = {Journal of the Acoustical Society of America},
    author = {Prosperetti, A. and Crum, L. A. and Commander, K. W.},
    number = {2},
    pages = {502--514},
    volume = {83},
    doi = {10.1121/1.396145}
}

@article{Supponen2016ScalingBubbles,
    title = {{Scaling laws for jets of single cavitation bubbles}},
    year = {2016},
    journal = {Journal of Fluid Mechanics},
    author = {Supponen, O. and Obreschkow, D. and Tinguely, M. and Kobel, P. and Dorsaz, N. and Farhat, M.},
    pages = {263--293},
    volume = {802},
    doi = {10.1017/jfm.2016.463}
}

@article{Brenner2002Single-bubbleSonoluminescence,
    title = {{Single-bubble sonoluminescence}},
    year = {2002},
    journal = {Reviews of Modern Physics},
    author = {Brenner, M. P. and Hilgenfeldt, S. and Lohse, D.},
    number = {2},
    pages = {425--484},
    volume = {74},
    doi = {10.1103/RevModPhys.74.425}
}

@article{Takemura2004TheNumber,
    title = {{The history force on a rapidly shrinking bubble rising at finite Reynolds number}},
    year = {2004},
    journal = {Physics of Fluids},
    author = {Takemura, F. and Magnaudet, J.},
    number = {9},
    pages = {3247--3255},
    volume = {16},
    doi = {10.1063/1.1760691}
}

@article{Magnaudet1998TheRadius,
    title = {{The viscous drag force on a spherical bubble with a time-dependent radius}},
    year = {1998},
    journal = {Physics of Fluids},
    author = {Magnaudet, J. and Legendre, D.},
    number = {3},
    volume = {10},
    doi = {10.1063/1.869582},
    issn = {10706631}
}

@book{Li2019TheoryDerivatives,
    title = {{Theory and Numerical Approximations of Fractional Integrals and Derivatives}},
    year = {2019},
    booktitle = {Theory and Numerical Approximations of Fractional Integrals and Derivatives},
    author = {Li, C. and Cai, M.},
    publisher = {Society for Industrial and Applied Mathematics},
    doi = {10.1137/1.9781611975888}
}

@article{Ferrara2007UltrasoundDelivery,
    title = {{Ultrasound microbubble contrast agents: Fundamentals and application to gene and drug delivery}},
    year = {2007},
    journal = {Annual Review of Biomedical Engineering},
    author = {Ferrara, K. and Pollard, R. and Borden, M.},
    pages = {415--447},
    volume = {9},
    doi = {10.1146/annurev.bioeng.8.061505.095852}
}

@book{White1991ViscousFlow,
    title = {{Viscous fluid flow}},
    year = {1991},
    author = {White, F. M.},
    edition = {2. ed},
    publisher = {McGraw-Hill Medical},
    address = {New York (N.Y.)},
    isbn = {0-07-069712-4 978-0-07-069712-6},
    language = {English}
}

\end{document}